\documentclass[11pt]{article}
\pdfoutput=1 

\usepackage{jinstpub} 

\usepackage{lineno}
\usepackage{xspace}
\usepackage{multirow}
\usepackage{array} 
\newcommand{\nhits}{\ensuremath{N_{\text{hits}}}\xspace}
\newcommand{\cTau}{\ensuremath{c\tau}\xspace} 
\newcommand{\PS}{\ensuremath{\text{S}}\xspace}
\newcommand{\PH}{\ensuremath{\mathrm{H}}}
\newcommand{\Pqb}{\ensuremath{\mathrm{b}}}
\newcommand{\Paqb}{\ensuremath{\mathrm{\overline{b}}}}
\newcommand{\bbbar}{\ensuremath{\Pqb\Paqb}\xspace}
\newcommand{\GeVns}{\ensuremath{\text{Ge\hspace{-.08em}V}}\xspace}
\newcommand{\pt}{\ensuremath{p_{\mathrm{T}}}\xspace}

\title{\boldmath High Multiplicity Trigger for Long-Lived Particles in CMS detector}

\author[a]{M.~Abbott}
\author[b]{D.~Acosta}
\author[c]{D.~Aebi}
\author[c]{M.~Ahmad}
\author[d]{A.~Albert}
\author[e]{A.~Apresyan}
\author[b]{C.~Arbour}
\author[f]{C.~Aruta}
\author[b]{K.~Banicz}
\author[f]{V.~Barashko}
% ...

\affiliation[a]{University of California, Davis, Davis, USA}

\affiliation[b]{Rice University, Houston, USA}

\affiliation[c]{Texas A\&M University, College Station, USA}

\affiliation[d]{California Institute of Technology, Pasadena, USA}

\affiliation[e]{Fermi National Accelerator Laboratory, Batavia, USA}

\affiliation[f]{University of Florida, Gainesville, USA}

% ...

\abstract{
Searches for long-lived particles (LLPs) at the CMS experiment often involve unconventional event topologies that are difficult to efficiently select using standard trigger strategies. To improve sensitivity to such signatures during LHC Run~3 operation, a dedicated High Multiplicity Trigger (HMT) has been developed and deployed in the CMS trigger system. The trigger targets events containing unusually large numbers of hits in the CMS cathode strip chamber (CSC) muon detectors, a characteristic signature of several LLP scenarios involving displaced decays in the muon system. The HMT implementation, trigger logic, rate dependence with pileup, and operational stability are described. Optimized hit multiplicity thresholds are used to maintain acceptable trigger rates under high-luminosity and high-pileup conditions while preserving high efficiency across a broad range of LLP lifetimes and kinematic regimes. The trigger performance is evaluated using both simulated event samples and proton-proton collision data collected during Run~3 of the LHC. The HMT substantially extends the CMS sensitivity to non-standard signatures associated with LLP decays and provides a flexible platform for future searches for physics beyond the Standard Model.
}


\begin{document}
\maketitle

\flushbottom

\section{Introduction}

The Standard Model (SM) of elementary particles, while providing the most accurate description of the microscopic world, is incomplete.
Many unexplained phenomena such as dark matter, dark energy, neutrino masses and baryogenesis hint at physics beyond the SM (BSM). 
Searches for BSM physics in proton-proton collisions at the Large Hadron Collider (LHC) at CERN cover a variety of different signatures and model assumptions, many of which look for signatures of stable or promptly decaying particles. 
However, there are many well motivated BSM models that produce new particles with macroscopic lifetimes and decay lengths comparable to the size of the detector. 
Such long-lived particles (LLPs) would produce unconventional detector signals that are challenging for online trigger and data processing systems deployed by the LHC experiments. 
Therefore, novel approaches are necessary to trigger on LLP signatures efficiently.

When LLPs decay inside the CMS muon system, the decay products can initiate electromagnetic or hadronic showers as the LLP traverses the steel that interweave the muon chambers in the CMS muon system.
The particle showers can create a large number of charged particle hits in the muon chambers, which is a distinctive signature that can be used to search for LLPs produced at the LHC.
Because this signature, called a Muon Detector Shower~(MDS), only makes very generic assumptions about the LLP decays, it is sensitive to a broad range of models beyond the SM, including exotic Higgs decays~\cite{Curtin_2015,Cheng_2016}, Heavy Neutral Leptons (HNL)~\cite{Cottin_2023}, axion-like particles~\cite{PhysRevLett.123.031803}, inelastic dark matter~\cite{Duerr_2021}, and hidden valley models~\cite{Strassler:2006im,Strassler:2006ri,Han:2007ae}.
However, all published CMS LLP searches using the MDS~\cite{cmscollaboration2024search_mds,cmscollaboration2024search_hnl,EXO20015,EXO23015,EXO24004} are severely trigger limited due to the lack of a dedicated trigger for MDS signatures.
Searches previously relied on large \pt imbalance, which typically has low signal efficiencies, or on associated electrons or muons, which limits the search channels. 
A High Multiplicity Trigger (HMT) dedicated for the MDS signature enables more comprehensive and sensitive searches for LLP models.

This paper reports on the first implementation of the HMT in both the Level-1 and the high-level triggers for LLPs decaying in the CMS endcap muon system during LHC Run-3. The initial implementation of the trigger was introduced to CMS data taking in 2022 with gradual improvements applied throughout Run-3. The paper is structured as follows. Section~\ref{sec:detector} provides an overview of the CMS detector, emphasizing the trigger system. 
The algorithm and its optimization are described in Section~\ref{sec:algo}, while the implementation is presented in Section~\ref{sec:implementation}. 
Section~\ref{sec:performance} describes the performance of the trigger, with results previously published by CMS in Refs.~\cite{EXO-23-016,CMS-DP-2024-099}, followed by future developments and a quick summary in Sections~\ref{sec:future} and~\ref{sec:summary}. 
 
\section{CMS Detector\label{sec:detector}}

The primary component of CMS is a superconducting solenoid with a 6~m internal diameter, which generates a magnetic field of 3.8~T. 
Various detectors are located within the solenoid, including a silicon pixel and strip tracker, an electromagnetic calorimeter (ECAL) made of lead tungstate crystal, and a hadron calorimeter (HCAL) made of brass and scintillators. 
These subdetectors are split into barrel and endcap sections, with forward calorimeters extending the pseudorapidity coverage ($\eta$). 
Muons are detected through gas-ionization chambers within the steel flux-return yoke outside the solenoid. 
For a complete description of the CMS detector, including coordinate system definitions and relevant kinematic variables, see Ref.~\cite{Chatrchyan:2008zzk}.

\begin{figure}[!ht]
  \centering
  \includegraphics[width=0.8\textwidth]{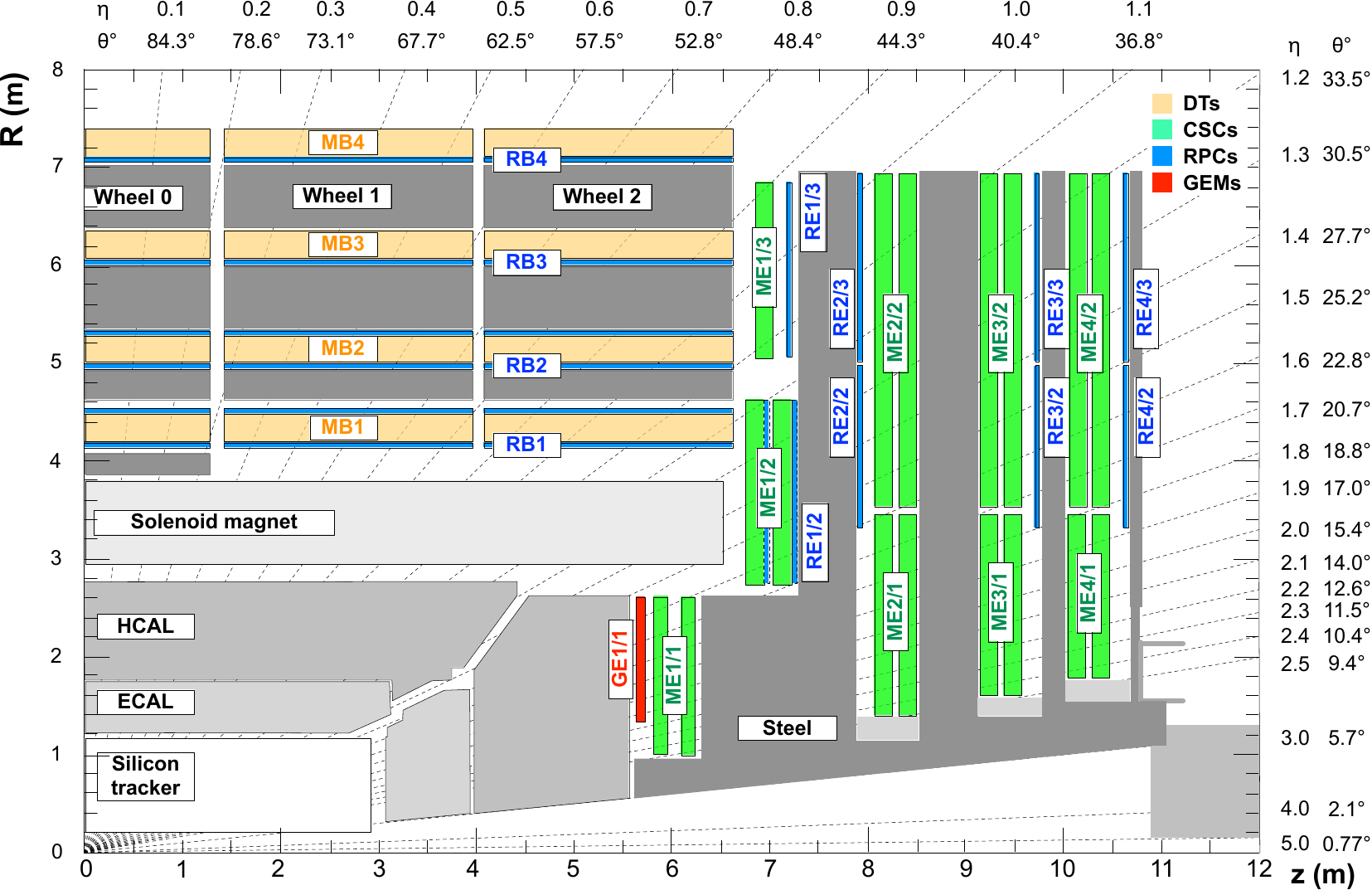}
  \caption{An \textit{R}-\textit{z} cross-section of a quadrant of the CMS detector with the axis parallel to the beam (\textit{z}) running horizontally and the radius (\textit{R}) increasing upward. The interaction point is at the lower left corner. 
The locations of the various muon stations and the steel flux-return disks (dark areas) are shown. The drift tube stations (DTs) are labeled MB (Muon Barrel), and the cathode strip chambers (CSCs) are labeled ME (Muon Endcap). Resistive plate chambers (RPCs) are mounted in both the barrel and endcaps of CMS, labeled RB and RE, respectively. Gas electron multiplier chambers (GEMs) are labeled GE. Image taken from~\cite{CMS:2023gfb}.}
  \label{fig:cms}
\end{figure}

The CMS muon system comprises four types of gas-ionization detectors: drift tubes (DTs), cathode strip chambers (CSCs), resistive-plate chambers (RPCs) and gas electron multipliers (GEMs).
First complete station of GEM chambers was installed ahead of Run-3 to better handle the increased particle rate expected at the HL-LHC.
Detailed descriptions of these chambers, including gas composition and operating voltage, can be found in Ref.~\cite{muon_performance,CMS:2023gfb}.
A schematic diagram of the CMS detector is shown in Figure~\ref{fig:cms}.

Due to bandwidth constraints, the HMT can only be implemented in CSCs during Run-3. The CSCs, situated in the pseudorapidity region $0.9 < \lvert\eta\rvert < 2.4$, function as standard multi-wire proportional chambers with a finely segmented cathode strip readout, facilitating precise measurement of the muon's position in the bending plane ($R$-$\phi$) as it traverses the gas volume. 
These chambers are located at four stations separated by the steel absorber of the magnet return yoke.
This arrangement resembles that of a sampling calorimeter, enabling the muon detectors to be used to identify displaced showers from the decay products of LLPs. 
This design also provides excellent background suppression when searching for LLP particles decaying in the muon detectors, due to the shielding material in front of and between stations.
The steel absorber, together with the calorimeters in front of the muon stations, provide a range of 12 to 27 interaction lengths of material between the first and the last muon stations~\cite{muon_performance}.

To identify collision events of potential interest for physics analyses, the CMS trigger system employs two levels of processing. The first level (L1) is implemented in custom-designed electronics and the High-Level Trigger (HLT) is executed on standard computers. 
The HLT further refines the selection of particle candidates and physics quantities reconstructed by L1, with an input event rate capped at 110~kHz during Run-3, targeting an average output rate of a few kHz for standard proton-proton events for offline storage and prompt reconstruction.

The L1 trigger utilizes information gathered from muon and calorimeter detectors, providing initial selections based on \textit{trigger primitives} (TPs), corresponding to coarse granularity and precision information. A set of algorithms is run over the events based on predefined criteria, constituting the so-called \textit{trigger menu}. 
Events meeting the conditions of at least one trigger in the menu, referred to as an L1 seed, proceed for further processing in the trigger chain. 
If that happens, the entire event readout from the data acquisition system is transmitted to the HLT. 
The assortment of algorithms in the menu reflects the diverse research interests within the CMS Collaboration. 
The L1 menu evolves in response to changing priorities in CMS physics and adjusts to beam or detector performance variations.

The L1 muon trigger system, described in Ref.~\cite{L1Phase1UpgradeTDR}, gathers inputs from all available subdetectors and is divided into three regional track reconstruction systems: the Barrel Muon Track Finder (BMTF), the Overlap Muon Track Finder (OMTF), and the Endcap Muon Track Finder (EMTF).
In particular, the HMT information is transmitted from the CSCs and received by the EMTF.
These track finders receive TP hits or short track segments from each station within their respective regions. 
They contain position ($\theta$ and $\phi$) coordinates, direction, and timing information correlated with the crossing of proton bunches at the LHC. 
Subsequently, muon reconstruction occurs on processor boards with field-programmable gate arrays (FPGAs) in each track finder.
The Global Muon Trigger collates reconstructed tracks from the track finders, eliminating geometrically overlapping tracks, before passing the remaining candidates to the L1 Global Trigger, which determines whether to retain the event for processing by the HLT. 
A detailed description of the L1 muon reconstruction can be found in dedicated publications~\cite{Khachatryan:2016bia,TRG-17-001}.

The HLT hardware comprises a cluster of multi-core servers, known as the Event Filter Farm, running a Linux operating system. 
Data processing within the HLT relies on the concept of the \textit{HLT paths}, which are sequences of algorithmic processing steps executed in a predefined order to reconstruct physics objects, such as leptons and QCD jets, calculate event level quantities, and make selections based on physics requirements. 
Specific L1 triggers (\textit{L1 seeds}) must be triggered before the HLT path execution begins. 
The reconstruction modules and selection filters within the HLT employ the same CMSSW~\cite{Jones:2015soc} software framework utilized for offline reconstruction and analyses, supporting multi-threaded event processing to optimize memory usage, a feature also used for the HLT software.
 
\section{Algorithm design\label{sec:algo}}

The defining feature of the LLP decays within the CMS muon system is the multiplicity of hits created by particle showers.
This starkly contrasts with what a typical muon would deposit in the detector, which are several hits in each detection layer that form a trajectory pointing to the interaction point. 
The multiplicity difference allows the design of simple yet highly efficient threshold algorithm to trigger on the potential LLP signals.
While the details of the particle shower depend on the kinematics of the LLP signal, a simple requirement on the minimum hit multiplicity is sufficient to discriminate against muons and other background events. 
The optimization of such thresholds will be discussed in detail in Section~\ref{sec:optimization}.
The simplicity of the algorithm is particularly important for the L1 system, which is subjected to relatively tight constraints in terms of bandwidth, latency, rates, and FPGA resources. 
With looser constraints at the HLT level, the goal of the algorithm is to create cluster objects that serve as the proxy of the LLP particle, being as close as possible to the ones used in the offline analysis, and make selections based on these cluster objects.  

Within the CSC endcap muon trigger system, adequate bandwidth was available to accommodate the HMT trigger primitives in the L1 trigger.
In other elements of the muon system, its implementation will be made possible after the HL-LHC upgrades which will provide the necessary resources.
Given the bandwidth constraint, encoding the hit multiplicity information is thus the most important goal of the algorithm at L1, while the details of the LLP shower are computed at the HLT. 
 
\section{Implementation\label{sec:implementation}}
\subsection{Implementation of the High Multiplicity Trigger in CSCs}
\begin{figure}
    \centering
    \includegraphics[width=0.8\textwidth]{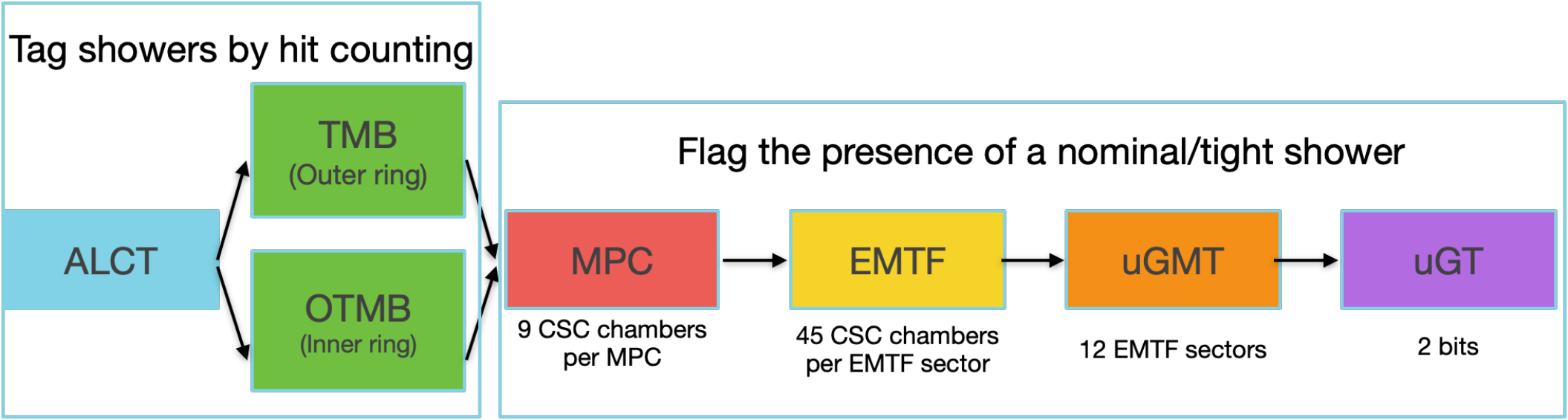}
    \caption{Illustration of the trigger path of HMT from CSCs through various stages of L1 trigger system. Dedicated new FPGA firmware is developed in ALCT, (O)TMB, MPC, EMTF, and uGMT to handle the additional information for the HMT trigger.}
    \label{fig:L1_flow}
\end{figure}

The overall strategy at L1 is to identify the LLP showers by counting the number of hits at individual chambers, then aggregate and condense the presence of showers from the chambers through the rest of the backend electronics.
Figure~\ref{fig:L1_flow} illustrates the flow of LLP information through the muon system trigger path.  

The fundamental quantities for building the LLP showers are the anode wire hits and cathode strip hits in each CSC chamber at each bunch crossing.
Anode wire hits are obtained by digitizing the analog current pulses from charges collected at the high-voltage wires in the CSC chamber. Cathode strip hits are obtained similarly from the induced current at the cathode, formed by comparing the induced charge on neighboring strips to locate the centroid at half-strip precision. Each CSC chamber consists of six layers, and each layer consists of 48-112 wires and 80-112 cathode strips, depending on the chamber’s location. 

A CSC chamber is identified as containing a shower when the number of anode or cathode hits exceeds configurable thresholds. The shower information is encoded using two bits corresponding to four categories: \emph{No Shower}, \emph{Loose}, \emph{Nominal}, and \emph{Tight}. Additional timing and layer-coincidence requirements are applied to suppress backgrounds and mitigate the effects of noisy channels that could otherwise generate high trigger rates.

In the outer-ring CSC chambers equipped with legacy Trigger Mother Boards (TMBs), only anode-based shower information is available. Only anode hits within the nominal bunch crossing (BX) are considered for the HMT. A valid anode HMT requires hits in at least five of the six CSC layers within the same BX. The HMT quantities are computed in the ALCT boards and encoded into two bits representing the shower category. The resulting HMT bits are transmitted to the EMTF through the TMB and Muon Port Card (MPC).

In the inner-ring CSC chambers equipped with Optical Trigger Mother Boards (OTMBs), both anode and cathode shower information are utilized. The OTMB receives the anode HMT bits from the ALCT board and independently computes the cathode HMT using cathode hits. Owing to the coarser timing resolution of cathode signals, cathode hits are evaluated within a three-BX window spanning BX$-1$, BX, and BX$+1$. A valid cathode HMT likewise requires hits in at least five of the six CSC layers. The OTMB then performs a timing-based matching of the anode and cathode HMT quantities within the allowed BX window and transmits the resulting 2-bit shower information to the EMTF through the MPC.



Each Endcap Muon Track Finder (EMTF) processor receives data from five MPCs and aggregates the shower information across the 45 input CSC chambers, which cover a sector of 60 degrees. The output bandwidth for shower information is limited to 4 bits per EMTF processor. 
Each EMTF processor performs a simple compression by checking the presence of at least 1 Loose/Nominal/Tight showers among the input chamber and encodes it into 2 bits. The remaining 2 bits are reserved for further developments. Additional protection against noisy chambers or faulty components is implemented in the EMTF processor, which ignores the input of any single CSC chamber that gives a persistently high rate within a short time interval. The Global Muon Trigger (GMT) receives inputs from all 12 EMTF sectors and checks for the presence of at least one Nominal or Tight shower in any sector. To increase the acceptance of LLP models that can produce more than one shower, GMT also checks for at least 2 Loose showers from different EMTF sectors. Finally, the Global Trigger (GT) encodes the information from the GMT into three L1 seeds (One-Nominal, One-Tight, and Two-Loose).

\subsection{Thresholds optimization\label{sec:optimization}}
Given the rate constraint, the anode and cathode hit thresholds in the CSC chambers are the main handles for optimizing signal efficiency.  
The CSC thresholds are optimized to maximize the signal efficiency while keeping a reasonable trigger rate at both L1 and HLT levels.
Zero-bias data collected in proton-proton collisions in 2022 were used to evaluate the expected rate. 
Monte Carlo (MC) samples were used to evaluate the signal efficiencies. The MC samples were created with the Pythia 8~\cite{bierlich2022comprehensive} event generator
using a benchmark model inspired by the twin Higgs scenario, in which the Standard Model-like Higgs boson decays into a pair of neutral long-lived scalars~\cite{OConnell:2006rsp},
each subsequently decaying into a pair of bottom quarks, $H \to 2S \to 4b$. The Higgs boson masses are scanned between 125 and 1000~GeV, LLP masses are scanned between
1 and 450 GeV, and displacements of the LLP in the particle frame are between 500~mm and 100~m. A total of 40000 events have been generated for each of them. 

\begin{figure}
\caption{Maximum number of counted cathode hits distributions for each CSC sector (the chamber with the maximum number of hits in a ring of chambers leads to the trigger decision). Merged 2022 and 2023 zero-bias data with emulated chamber hits are shown with solid black points. Filled templates show two different signal MC samples: $m_{H}$ = 125 GeV, $m_{S}$ = 12 GeV, $c\tau_{S}$ = 9000 mm, and $m_{H}$ = 350 GeV, $m_{S}$ = 80 GeV, $c\tau_{S}$ = 10000 mm. Vertical lines represent the Loose and Nominal thresholds set in each chamber. The distributions are normalized to unity.}
\label{fig:comp_max}
\centering
\includegraphics[width=0.33\textwidth]{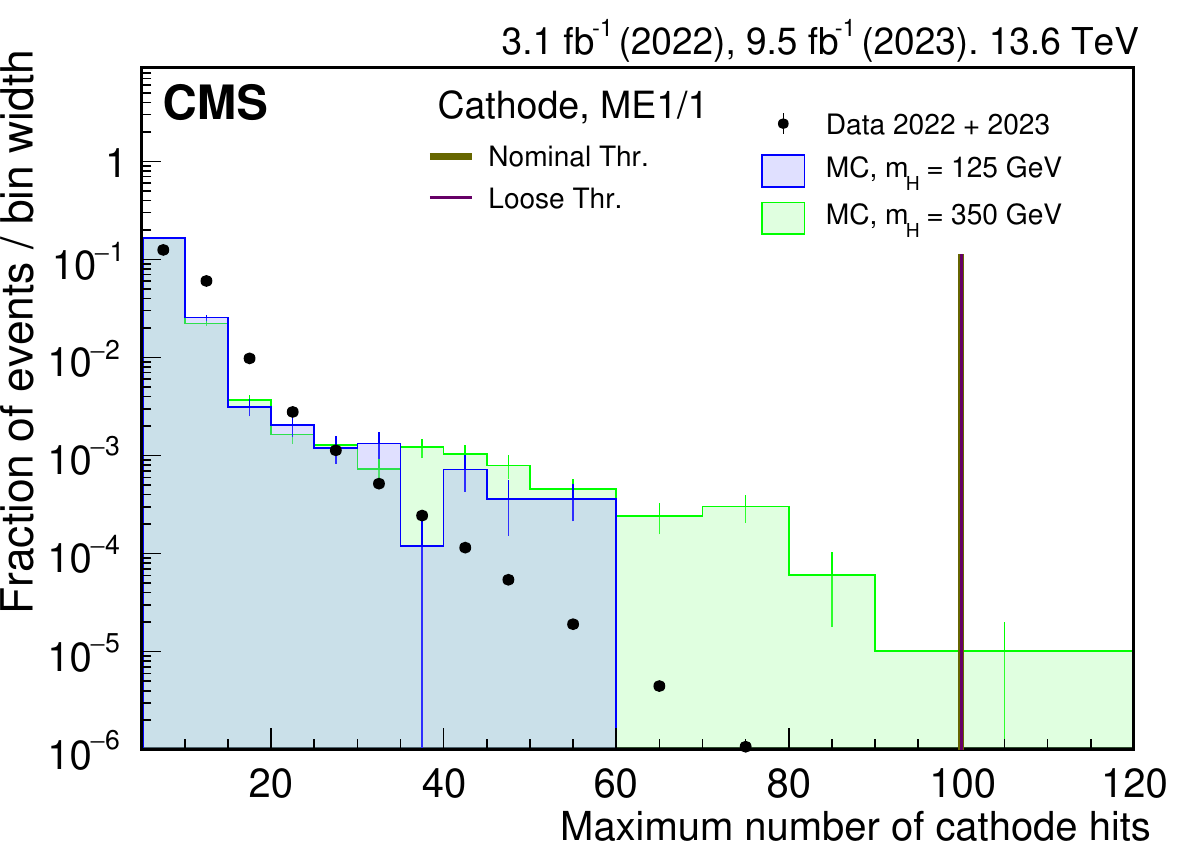}
\includegraphics[width=0.33\textwidth]{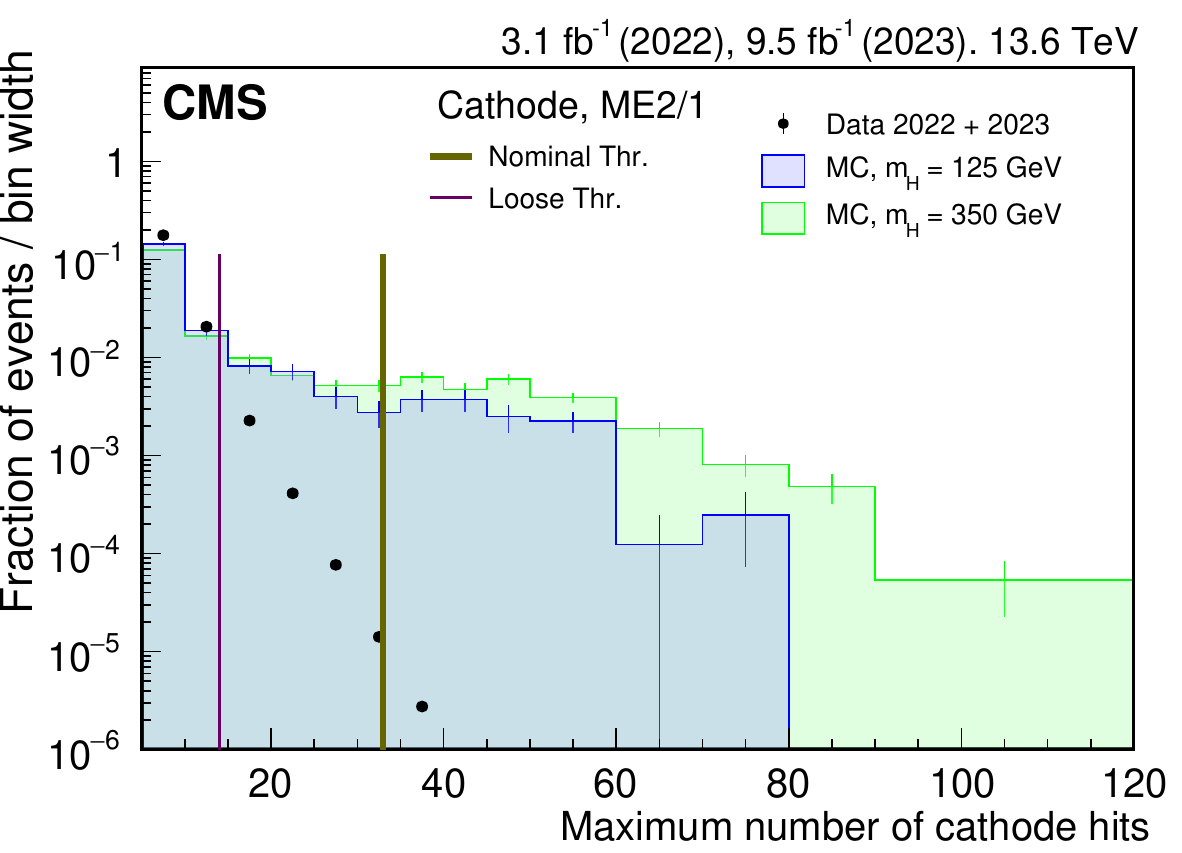}\\
\includegraphics[width=0.33\textwidth]{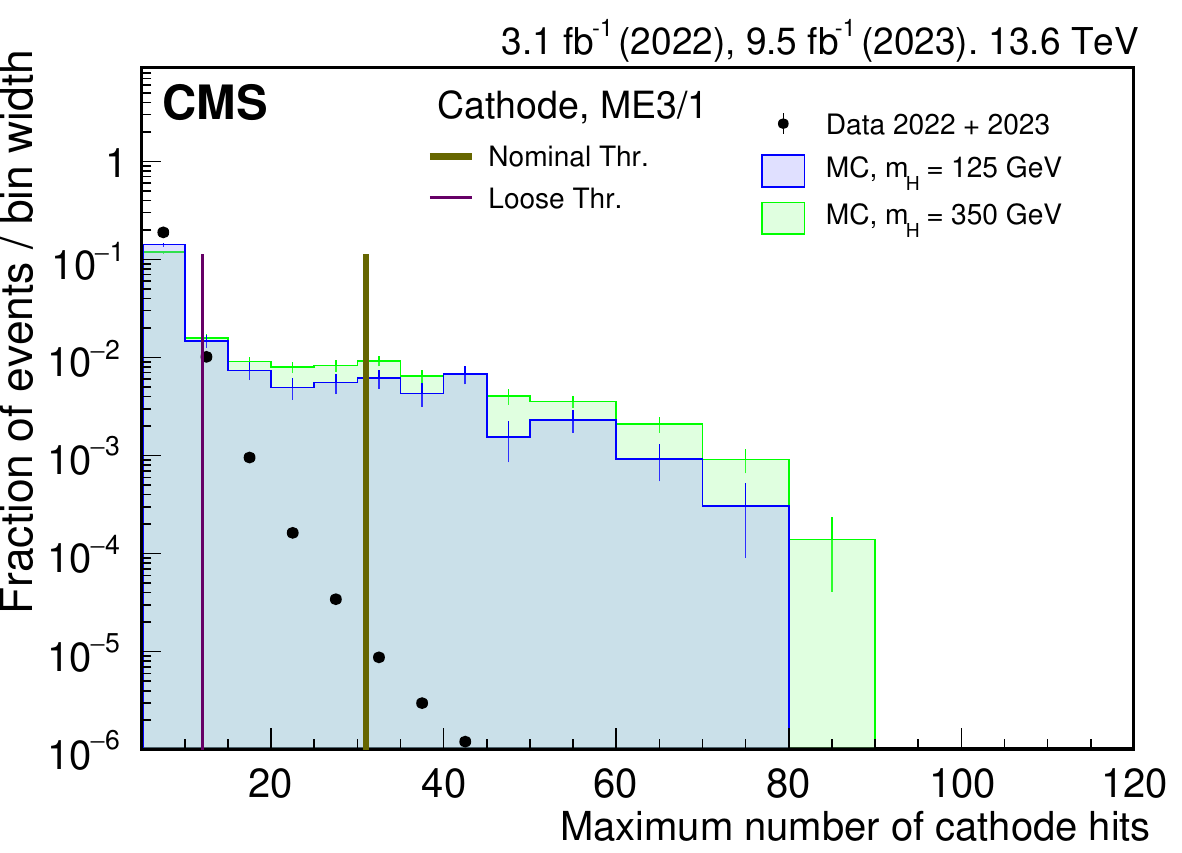}
\includegraphics[width=0.33\textwidth]{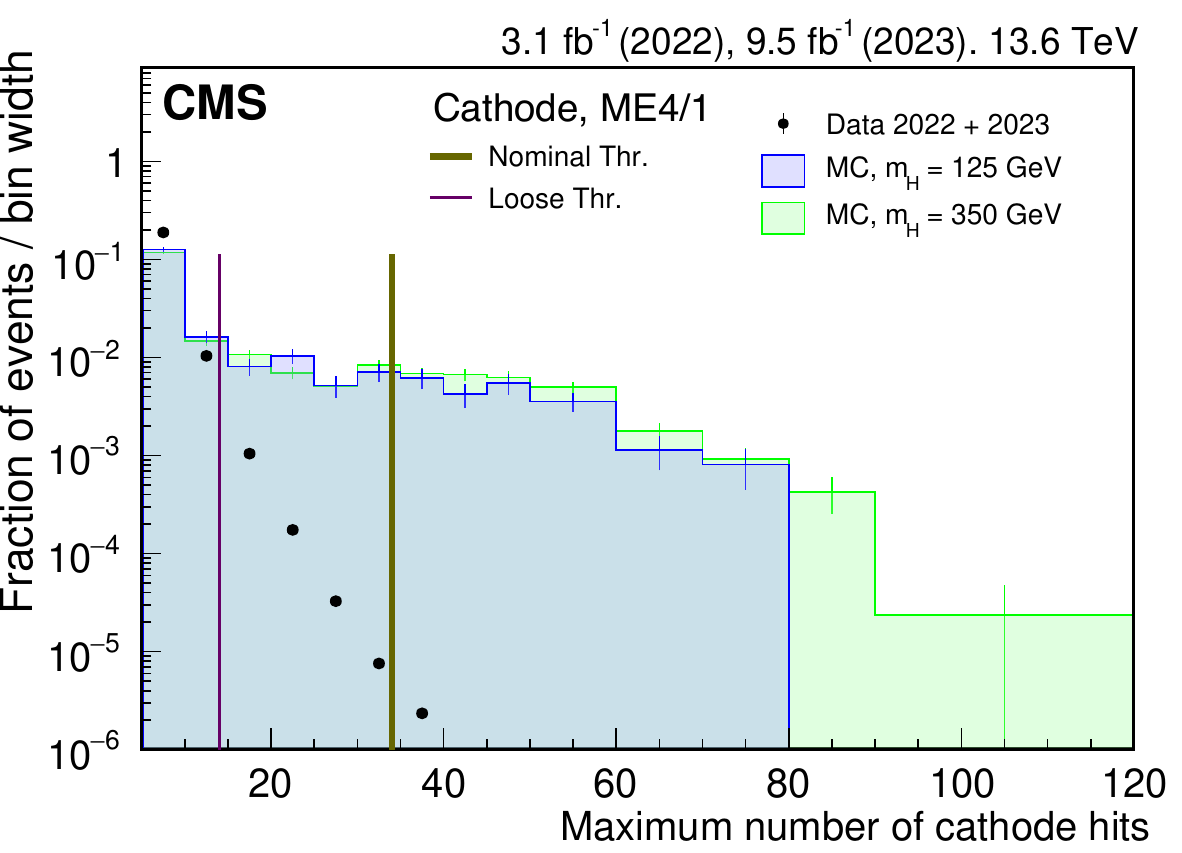}\\
\end{figure}
\begin{figure}
\caption{Maximum number of counted anode hits distributions for each CSC sector (the chamber with the maximum number of hits in a ring of chambers leads to the trigger decision). Merged 2022 and 2023 zero-bias data with emulated chamber hits are shown with solid black points. Filled templates show two different signal MC samples: $m_{H}$ = 125 GeV, $m_{S}$ = 12 GeV, $c\tau_{S}$ = 9000 mm, and $m_{H}$ = 350 GeV, $m_{S}$ = 80 GeV, $c\tau_{S}$ = 10000 mm. Vertical lines represent the Loose and Nominal thresholds set in each chamber. The distributions are normalized to unity.}
\label{fig:wire_max}
\centering
\includegraphics[width=0.33\textwidth]{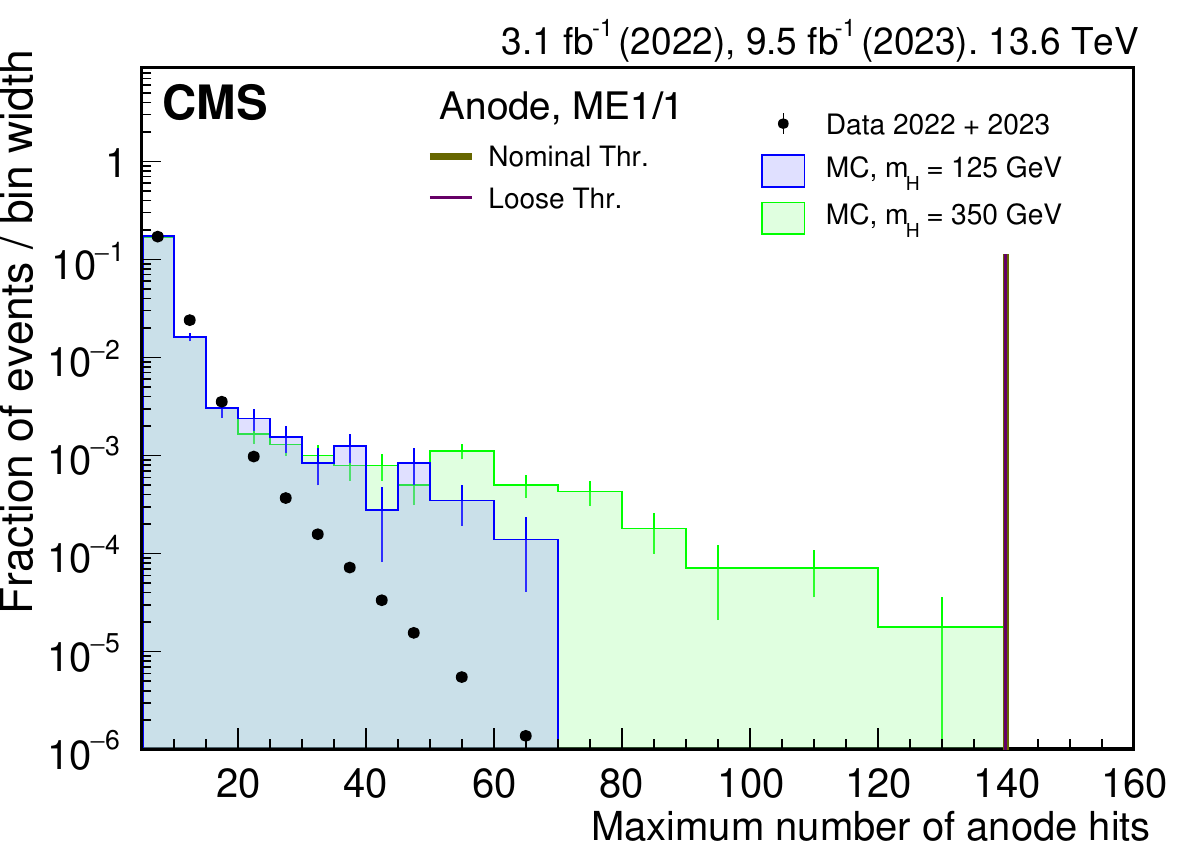}
\includegraphics[width=0.33\textwidth]{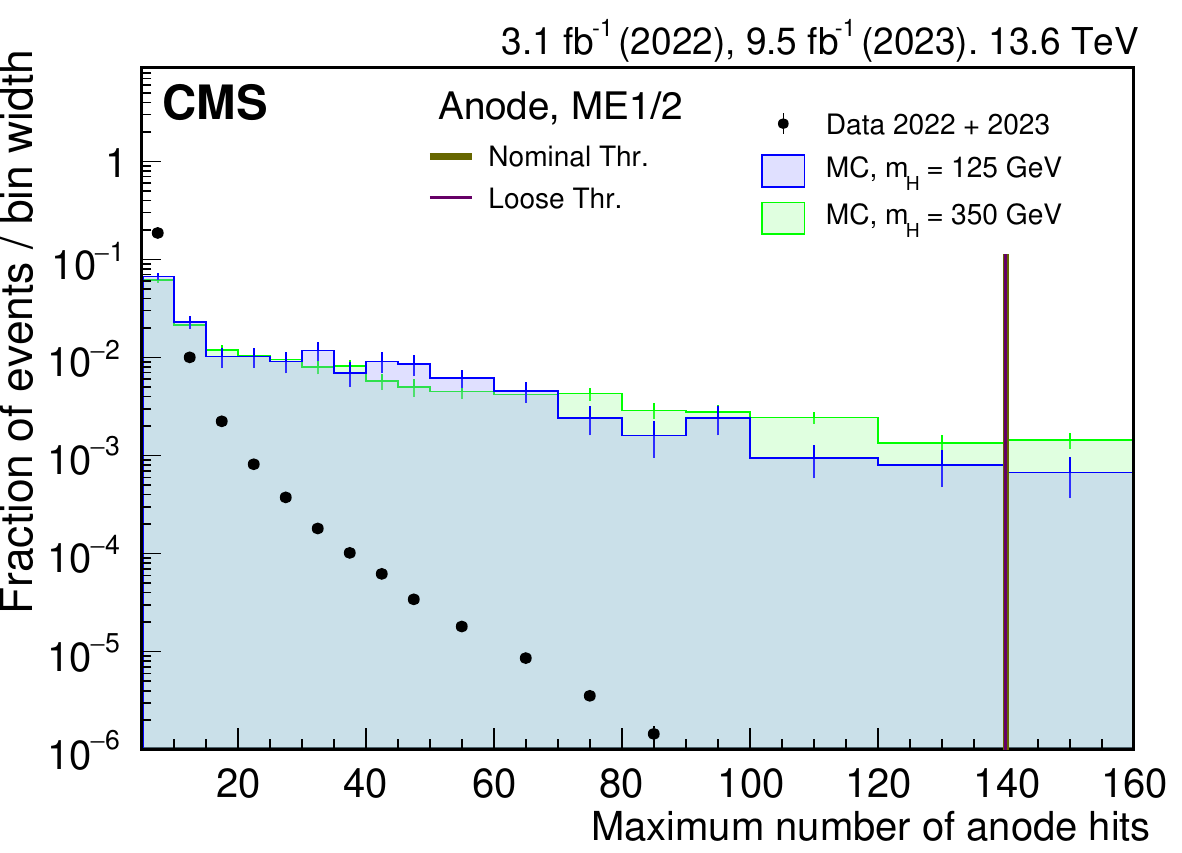}\\
\includegraphics[width=0.33\textwidth]{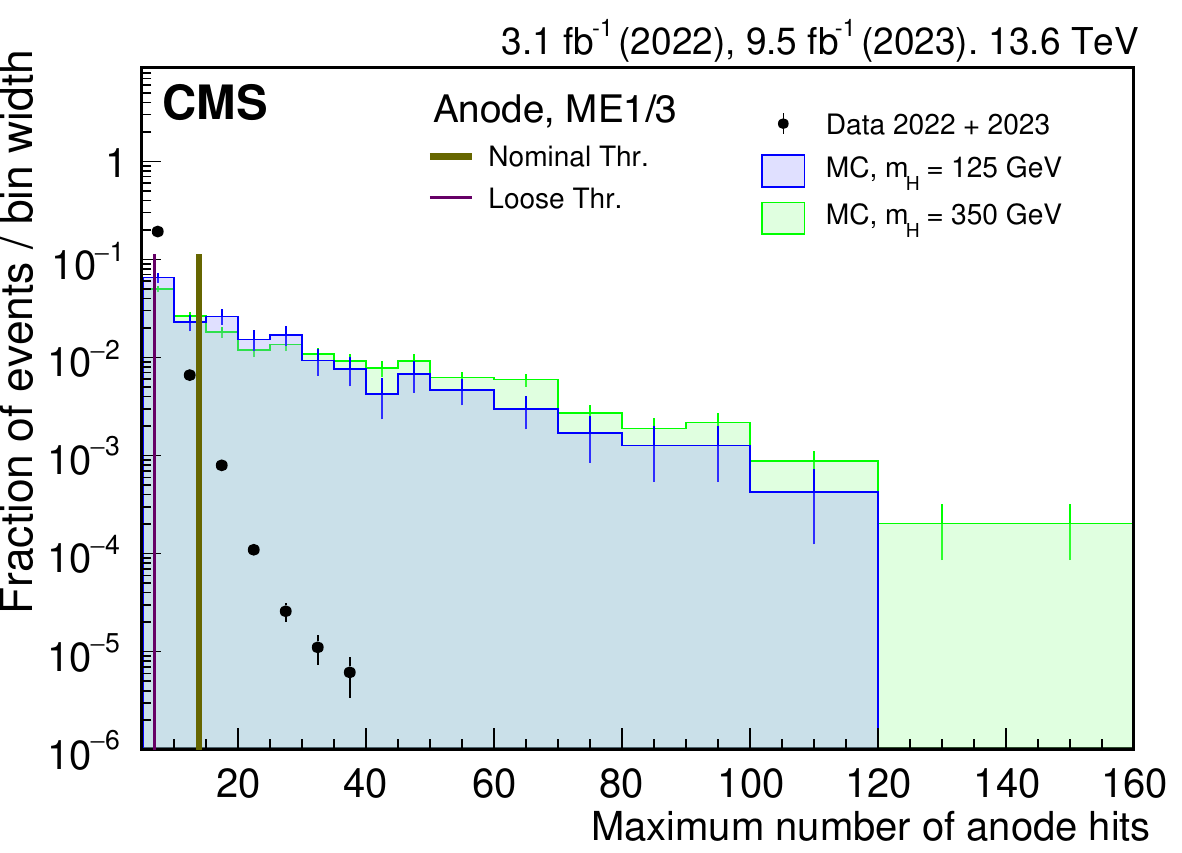}
\includegraphics[width=0.33\textwidth]{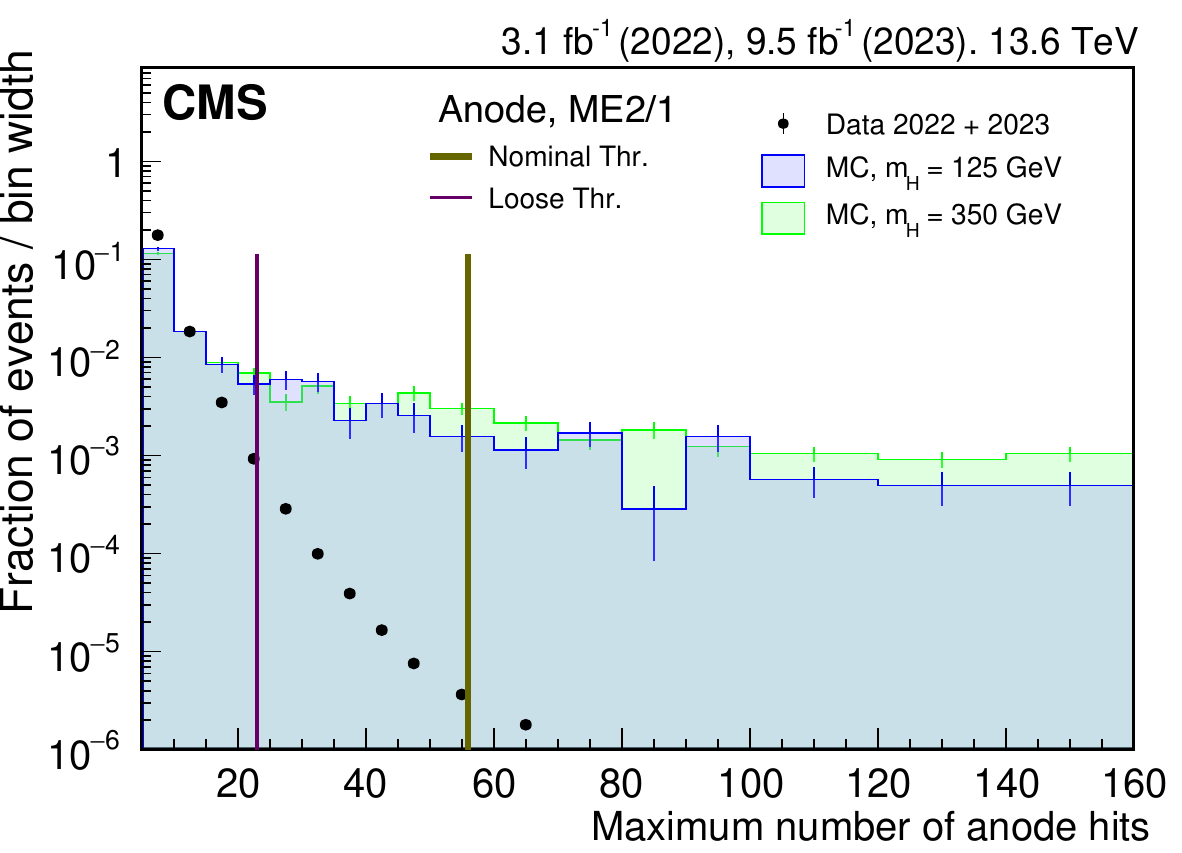}\\
\includegraphics[width=0.33\textwidth]{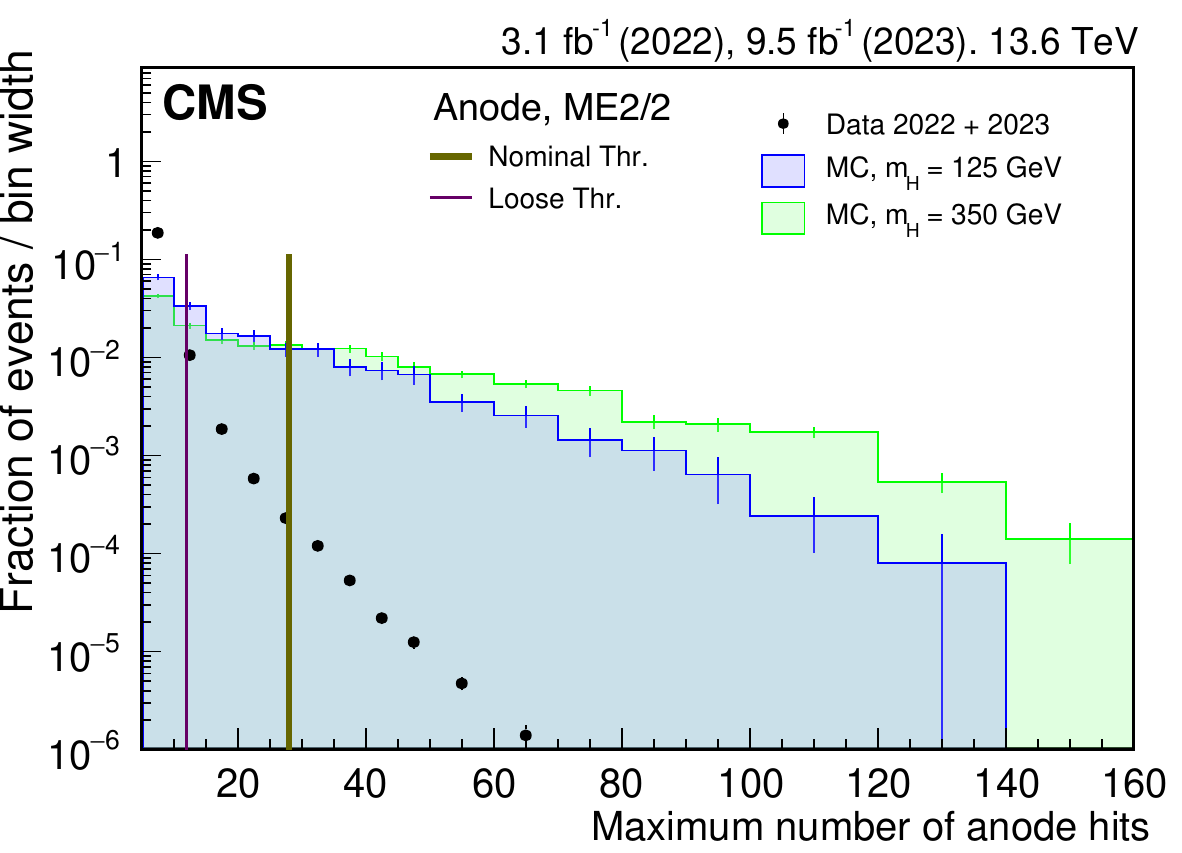}
\includegraphics[width=0.33\textwidth]{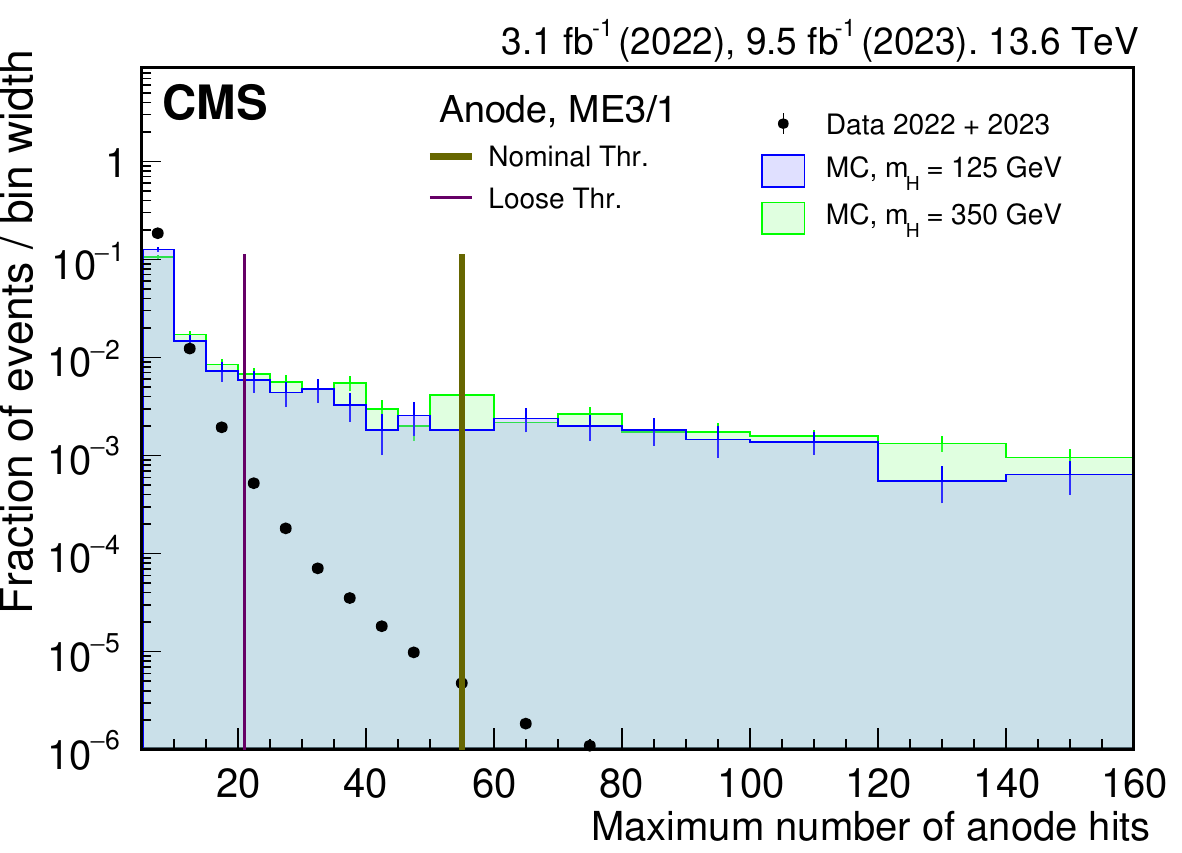}\\
\includegraphics[width=0.33\textwidth]{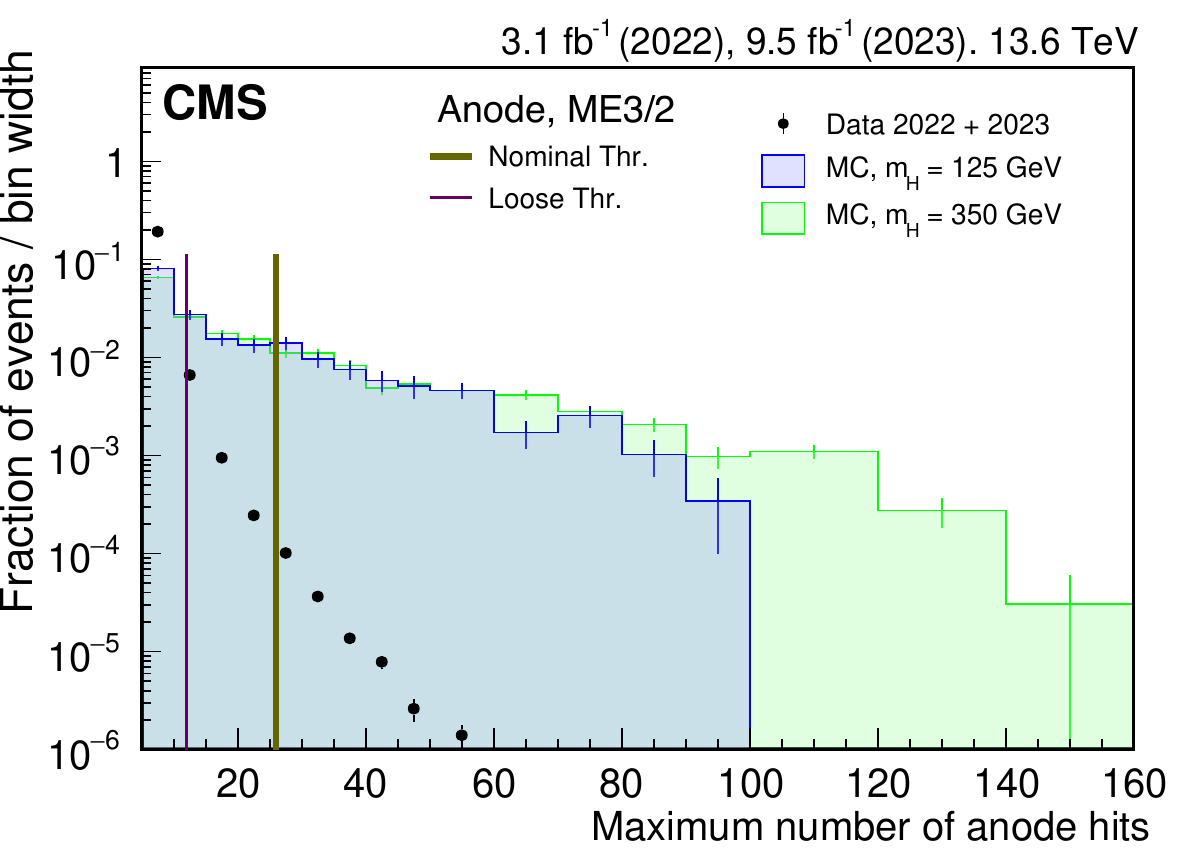}
\includegraphics[width=0.33\textwidth]{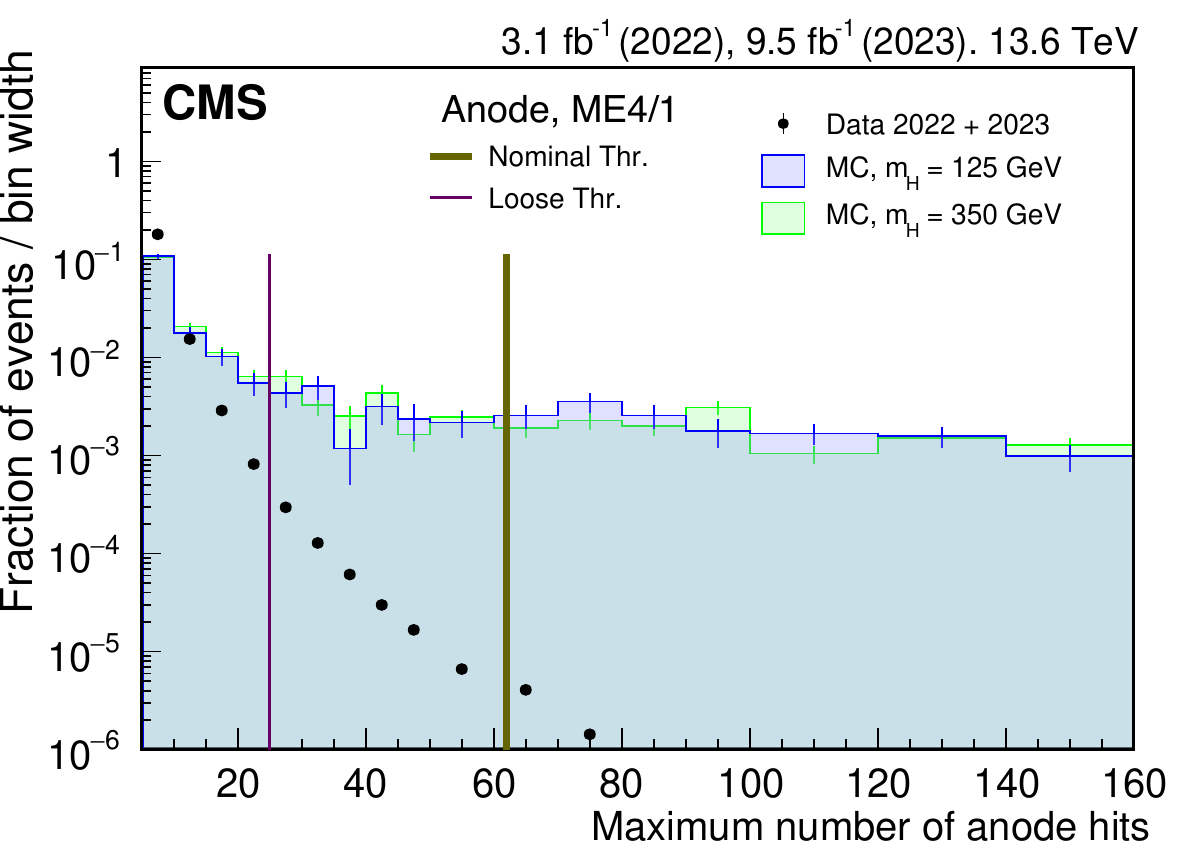}\\
\includegraphics[width=0.33\textwidth]{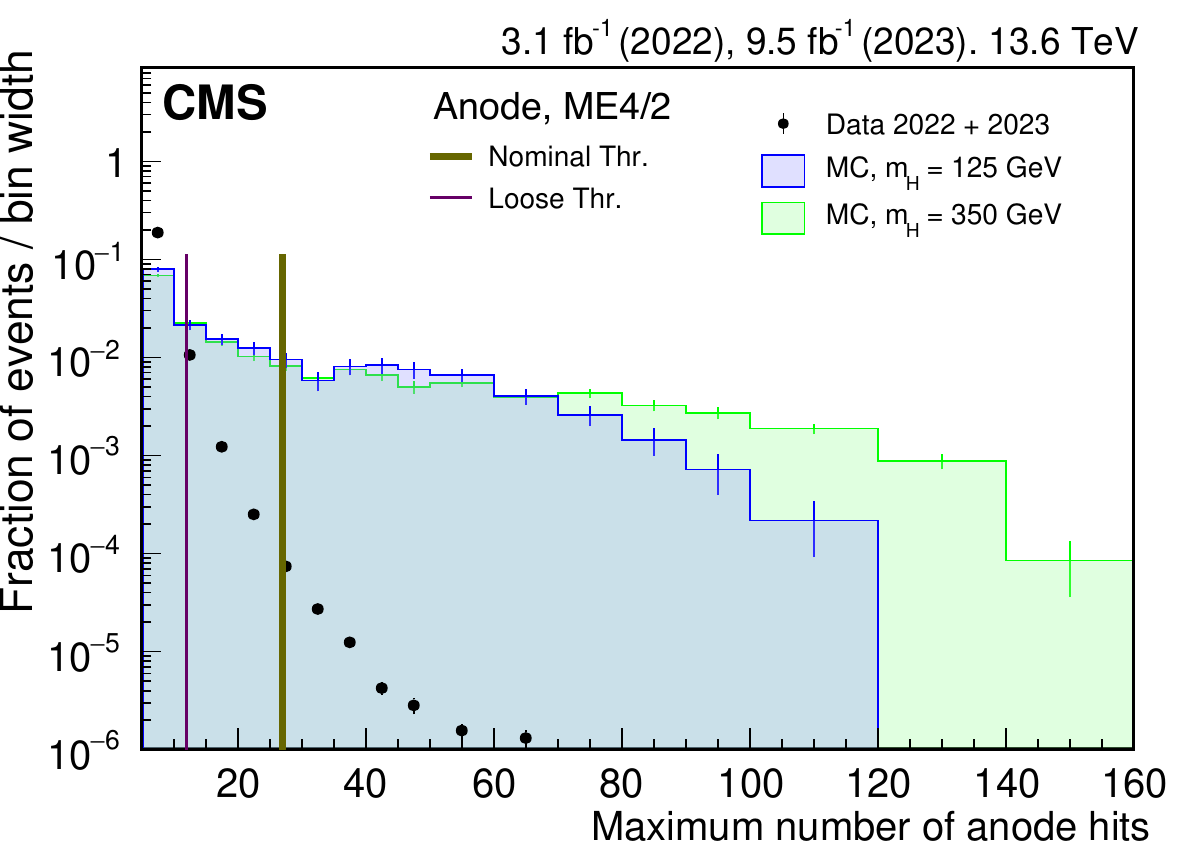}
\end{figure}

The chosen thresholds that showed a good compromise between signal efficiency and rate are listed in Tables~\ref{table:thres_ano} and~\ref{table:thres_cat}. 
The most stringent thresholds are set in the innermost chambers of the first station, ME1/1 and ME1/2, due to the larger background rate in these chambers caused by the lesser amount of steel between them and the collision point. Since the rate of background events is highly dependent on $\eta$, the thresholds chosen for chambers in the inner rings are higher to suppress the background. 
Distributions of the maximum number of cathode and anode hits in a CSC chamber, shown separately for each station-ring, are presented in Figures~\ref{fig:comp_max} and~\ref{fig:wire_max}.
It can be observed that the data distributions typically fall faster as the maximum number of hits increases.
The signal MC tends to fall slower and has a longer tail. 
The size of the tail is primarily driven by the LLP mass: the higher the particle mass, the more energy is expected to be deposited in the chamber, and the greater the penetrating power of the showering LLP.

\begin{table}[htbp]
  \centering
    \caption{Set thresholds on anode-wire counts for Loose, Nominal, and Tight showers.} \label{table:thres_ano}
      \begin{tabular} {cccc}
        \hline
        CSC station-ring & Loose Thr. & Nominal Thr. & Tight Thr. \\ 
        \hline
        ME1/1 & 140 & 140 & 140   \\
        ME1/2 & 140 & 140 & 140   \\
        ME1/3 &   7 &  14 & 18   \\
        ME2/1 &  23 &  56 & 58   \\
        ME2/2 &  12 &  28 & 32   \\
        ME3/1 &  21 &  55 & 57   \\
        ME3/2 &  12 &  26 & 34   \\
        ME4/1 &  25 &  62 & 64   \\
        ME4/2 &  12 &  27 & 31   \\
        \hline
      \end{tabular}
\end{table}

\begin{table}[htbp]
  \centering
    \caption{Set thresholds on cathode-strip counts for Loose, Nominal, and Tight showers.} \label{table:thres_cat}
      \begin{tabular} {cccc}
        \hline
        CSC station-ring & Loose Thr. & Nominal Thr. & Tight Thr. \\ 
        \hline
        ME1/1 & 100 & 100 & 100   \\
        ME2/1 & 14 & 33 & 35   \\
        ME3/1 & 12 & 31 & 33   \\
        ME4/1 & 14 & 34 & 36   \\
        \hline
      \end{tabular}
\end{table}

\clearpage
\subsection{Implementation at HLT}
With looser constraints on latency compared to L1, more shower properties can be reconstructed at the HLT level.
Contrary to the algorithm at L1, where the hit counting has to be done at a single CSC chamber due to bandwidth constraint, hits from multiple CSC chambers or stations are available at HLT, which are reconstructed as a cluster object.
This allows the HLT to capture the properties of signal showers more precisely, thus rejecting more backgrounds.

The first step in reconstructing the cluster object is to refine the input hits from L1.
The anode and cathode hits used in the L1 are reconstructed into hits with 2-D positions within a chamber, called a rechit, corresponding to the positions of a charged particle traversing the chamber. 
Together with the positions of the chambers, the CSC rechits are 3-D spatial points that can be used as input to cluster algorithms.
We choose to cluster the CSC rechits using the Cambridge-Aachen~(CA) algorithm~\cite{Dokshitzer_1997, wobisch1999hadronization} in the $\eta-\phi$ space with a distance parameter of 0.4, which is chosen to give clusters of similar radius as the ones used in the offline analysis~\cite{cmscollaboration2024search_hnl,cmscollaboration2024search_mds}.
A cluster object is defined to have at least 50 rechits, a threshold twice as high as a typical muon would create in the muon system. 
Hereafter, \nhits denotes the number of rechits in a cluster.
Cluster properties are then computed based on the constituent rechits.

Based on these cluster objects, many HLT selections can be designed to target specific LLP models, especially when combined with other physics objects such as leptons and photons.
Here, we describe the design of two new generic HLT paths, called the One-Nominal and the Two-Loose path, which are driven by the background rates with minimal assumptions on the signal model.

The One-Nominal HLT path is seeded by the One-Nominal L1 seed and reduces the event rate by a factor of approximately 170 by exploiting features of the background clusters.
Background clusters tend to be concentrated in a single station and have higher $\eta$, whereas signal clusters could spread across different stations. 
Therefore, we set separate thresholds of \nhits in the One-Nominal HLT path, depending on the cluster's $\eta$ and the number of stations with more than 10 rechits in the cluster. 
Table~\ref{table:OneNominalHLT} shows the optimized thresholds, ranging between 100 and 500 hits,  for the 4 categories in the One-Nominal HLT path. 

Timing information of the rechits can be used to suppress background clusters that are not in-time with the nominal bunch crossing.
The cluster time, $t_{cls}$, is defined as 
\begin{equation*}
    t_{cls} = \frac{\sum (t_{\text{wire}} + t_{\text{cathode}})/2 }{\nhits},
\end{equation*}
where $t_{\text{wire}}$ and $t_{\text{cathode}}$ are the times measured with anodes and cathodes of the constituent rechits, is required to be between $-5.0$ and $12.5$~ns.
Because there is the smallest amount of shielding materials for the ME1/1 and ME1/2 stations, the background rates are the highest at those stations.
Therefore, clusters passing the One-nominal path are required to have zero rechits in ME1/1 and ME1/2 stations.   

Similarly, the Two-Loose HLT path is seeded by events passing the One-Nominal or the Two-Loose L1 seeds and keeps events with at least two clusters with a minimum \nhits of 75 in each cluster. 
Requiring the presence of two clusters in the events reduces the rate drastically, so lower \nhits selections are allowed.
Time selection and the ME1/1 and ME1/2 vetoes are also required for the Two-Loose HLT path.

\begin{table}[htbp]
\centering
\caption{Selections on the cluster hit multiplicity for the HLT paths. The One-Nominal trigger requires different thresholds on the \nhits depending on the pseudorapidity of the cluster and its spread along the CSC stations. The Two-Loose trigger has a lower \nhits threshold because the requirement of an addition cluster heavily reduces trigger rate. The $\eta$ boundary corresponds to the approximate boundary between inner and outer CSC rings, which have different L1 thresholds.} 
\label{table:OneNominalHLT}        
\begin{tabular}{c|cc|c}
\hline
    & \multicolumn{2}{c|}{One-Nominal Trigger} & {Two-Loose Trigger} \\
\hline
L1 seeds & \multicolumn{2}{c|}{One-Nominal }  & One-Nominal OR Two-Loose \\
\hline
$\lvert\eta\rvert$     & $\lvert\eta\rvert \leq 1.9$ & $\lvert\eta\rvert > 1.9$ & no selection\\
\hline
Number of stations = 1            & 200  & 500  & \multirow{2}{*}{75}    \\
Number of stations \textgreater 1 & 100  & 500  &   \\
\hline
\end{tabular}
\end{table}

\section{Performance\label{sec:performance}}
Performance of the triggers described in this paper has been published by the CMS Collaboration in Refs.~\cite{EXO-23-016,CMS-DP-2024-099}.
The plots shown here are reproduced from that work and represent a subset of the results presented therein.

Signal efficiencies for the benchmark model are shown in Tab.~\ref{table:signal_eff_LLP} for MC events with at least one and two LLPs decaying in the CSC detector acceptance, respectively.
These efficiencies represent a dramatic improvement of more than 30 times compared to the Run-2 analysis~\cite{cmscollaboration2024search_mds}, where the same benchmark model was triggered using large missing transverse momentum with a signal efficiency of approximately 1\%, owing to the absence of a dedicated MDS trigger.
In the latter case with two LLPs decaying in the CSC detector acceptance, the presence of the Two-Loose L1 seed yields up to a 10\% increase in the signal efficiency, when compared to only using the One-Nominal seed. 

\begin{table}[tbp]
  \centering
    \caption{Signal efficiency $\epsilon$ in $\PH \to \PS\PS \to \bbbar\bbbar$ events, for different choices of $m_{\PH}$, $m_{\PS}$, and $\cTau_{\PS}$, for simulated events with at least 1 (middle columns) and exactly 2 (right columns) LLPs decaying within the CSC detector acceptance. The $\epsilon_{\text{acc}}$ refers to the percentage of events within the CSC detector acceptance. The $\epsilon_{\text{1-N}}$, $\epsilon_{\text{1-T}}$, and $\epsilon_{\text{2-L}}$ denote the efficiencies for the One-Nominal, One-Tight, and Two-Loose algorithms, respectively.} \label{table:signal_eff_LLP}
    \begin{tabular}{@{}%
        >{\centering\arraybackslash}p{1.35cm}%
        >{\centering\arraybackslash}p{1.35cm}%
        >{\centering\arraybackslash}p{1.35cm}|
        ccc|ccc@{}} \hline 
        \multirow{2}{*}{$m_{\PH}$ [{\GeVns}]} & 
        \multirow{2}{*}{$m_{\PS}$ [{\GeVns}]} & 
        \multirow{2}{*}{$\cTau_{\PS}$ [{mm}]} 
        & \multicolumn{3}{c|}{$\geq 1$ LLP decay in CSCs} 
        & \multicolumn{3}{c}{2 LLP decays in CSCs} \\
         &  &  
        & $\epsilon_{\text{acc}}$ & $\epsilon_{\text{1-N}}$ & $\epsilon_{\text{1-T}}$ 
        & $\epsilon_{\text{acc}}$ & $\epsilon_{\text{1-N}}$ & 
        $\epsilon_{\text{2-L}}$ OR $\epsilon_{\text{1-N}}$\\
        \hline
        125 & 12  & 900   & 16.2\% & 43.1\% & 39.2\% & 0.7\% & 69.6\% & 75.0\% \\
        125 & 25  & 1500  & 15.9\% & 43.7\% & 40.0\% & 0.7\% & 75.1\% & 79.7\% \\
        250 & 60  & 10000 & 10.3\% & 55.5\% & 52.1\% & 0.3\% & 74.0\% & 78.2\% \\
        250 & 120 & 10000 & 15.4\% & 29.9\% & 27.7\% & 1.0\% & 44.4\% & 49.6\% \\
        350 & 160 & 10000 & 17.9\% & 39.1\% & 36.7\% & 1.1\% & 63.4\% & 69.6\% \\
        \hline
    \end{tabular}
\end{table}

\begin{figure}[!hbtp]
\caption{Measured rate of the L1 seeds as a function of pileup for a 2022 data-taking run (left) and for a 2023 data-taking run (right). The datasets are fitted by linear functions, which are shown overlaid. Figures taken from Refs.~\cite{EXO-23-016,CMS-DP-2024-099}.}
\label{fig:rate_vs_pu}
\centering
\includegraphics[width=0.49\textwidth]{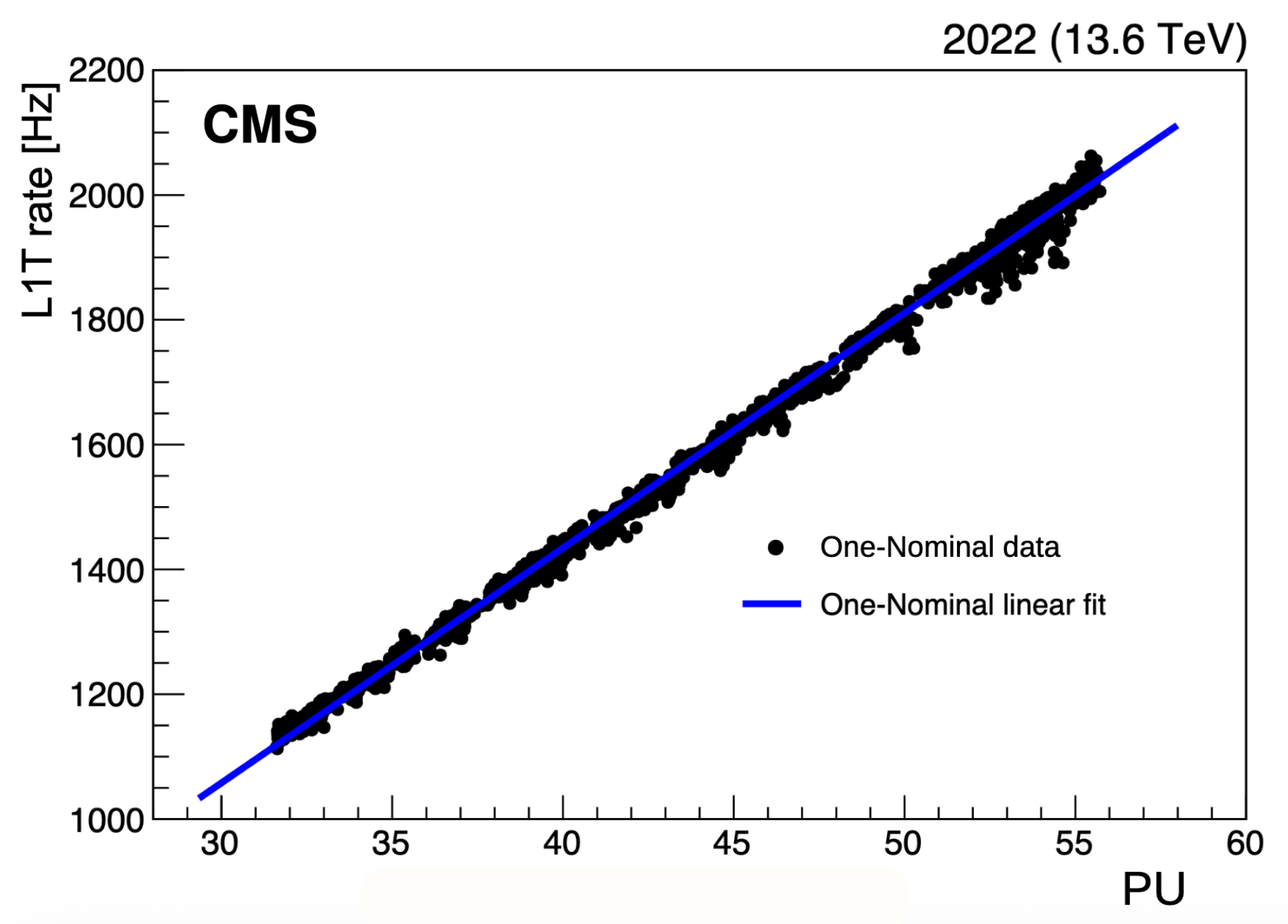}
\includegraphics[width=0.49\textwidth]{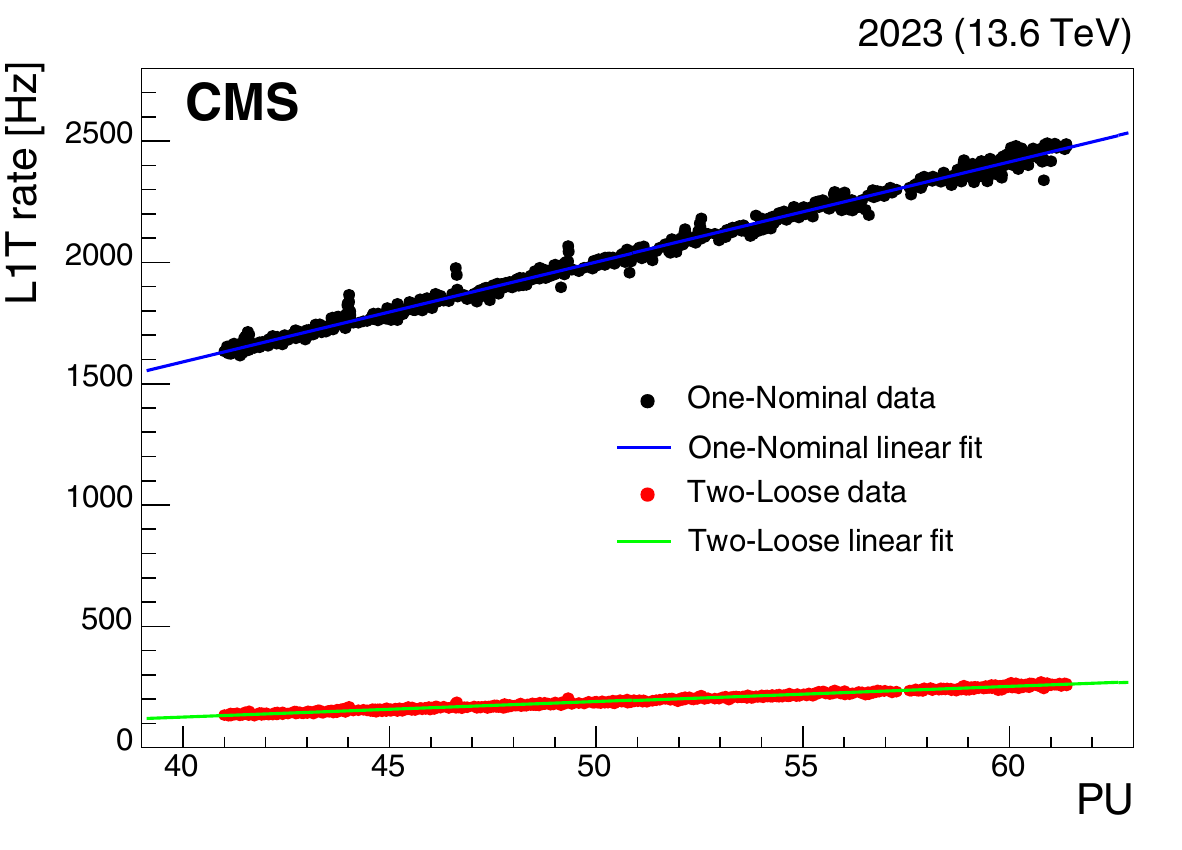}
\end{figure}

Figure~\ref{fig:rate_vs_pu} shows the linear dependence of the rate of the L1 seeds with the pileup for two data-taking runs in 2022 and 2023, demonstrating good stability of the triggers over different data-taking conditions at the LHC. 
The Two-Loose rate remains approximately $\sim10\%$ of the One-Nominal rate, illustrating that the backgrounds are heavily suppressed by the presence of an additional cluster object.
HMT triggers occupy approximately 1-2\% of the total CMS L1 trigger bandwidth in 2022-2023 data-taking conditions.
Similar to the L1 seeds, the rates at HLT maintain a linear dependence with pileup with a rate around 14~Hz at the pileup of 55, representing less than 0.5\% of the total CMS HLT output rate under typical Run-3 conditions.

\begin{figure}[!hbtp]
\caption{L1 trigger efficiencies of the One-Nominal trigger for outer ring chambers as a function of offline cluster size. Events are selected from the single-muon trigger and have exactly one cluster matched to a muon with $\Delta R <$ 0.4 , with at least 90\% of rechits in a single station-ring. \nhits refers to the number of offline CSC rechits after CA clustering with a distance parameter of 0.4. 
The turn-on feature is a result of the differences between the trigger object used in L1 and HLT (anode hit counts and HLT rechits respectively). Figures taken from Refs.~\cite{EXO-23-016,CMS-DP-2024-099}.}
\label{fig:outer_ring_eff}
\centering
\includegraphics[width=0.49\textwidth]{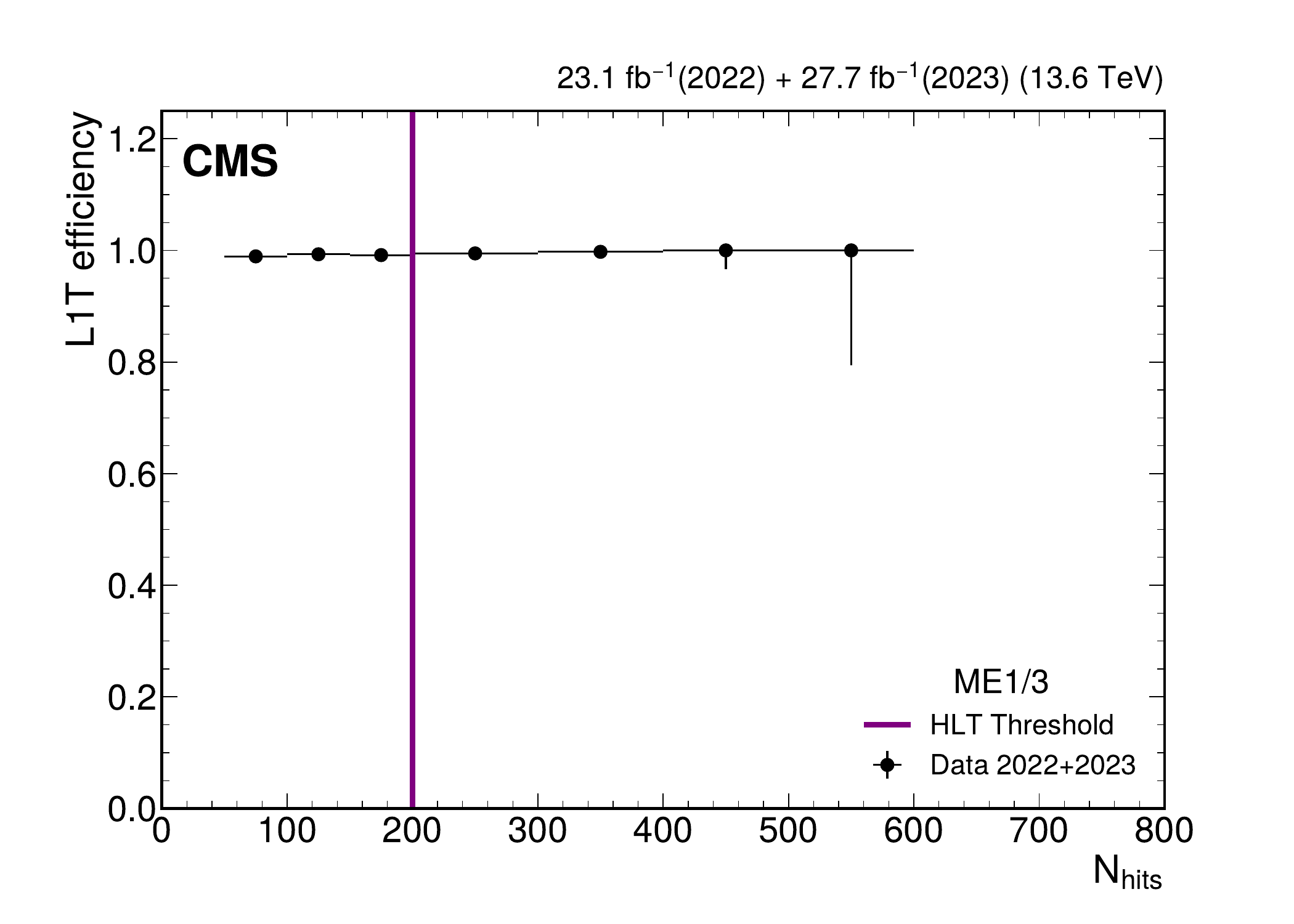}
\includegraphics[width=0.49\textwidth]{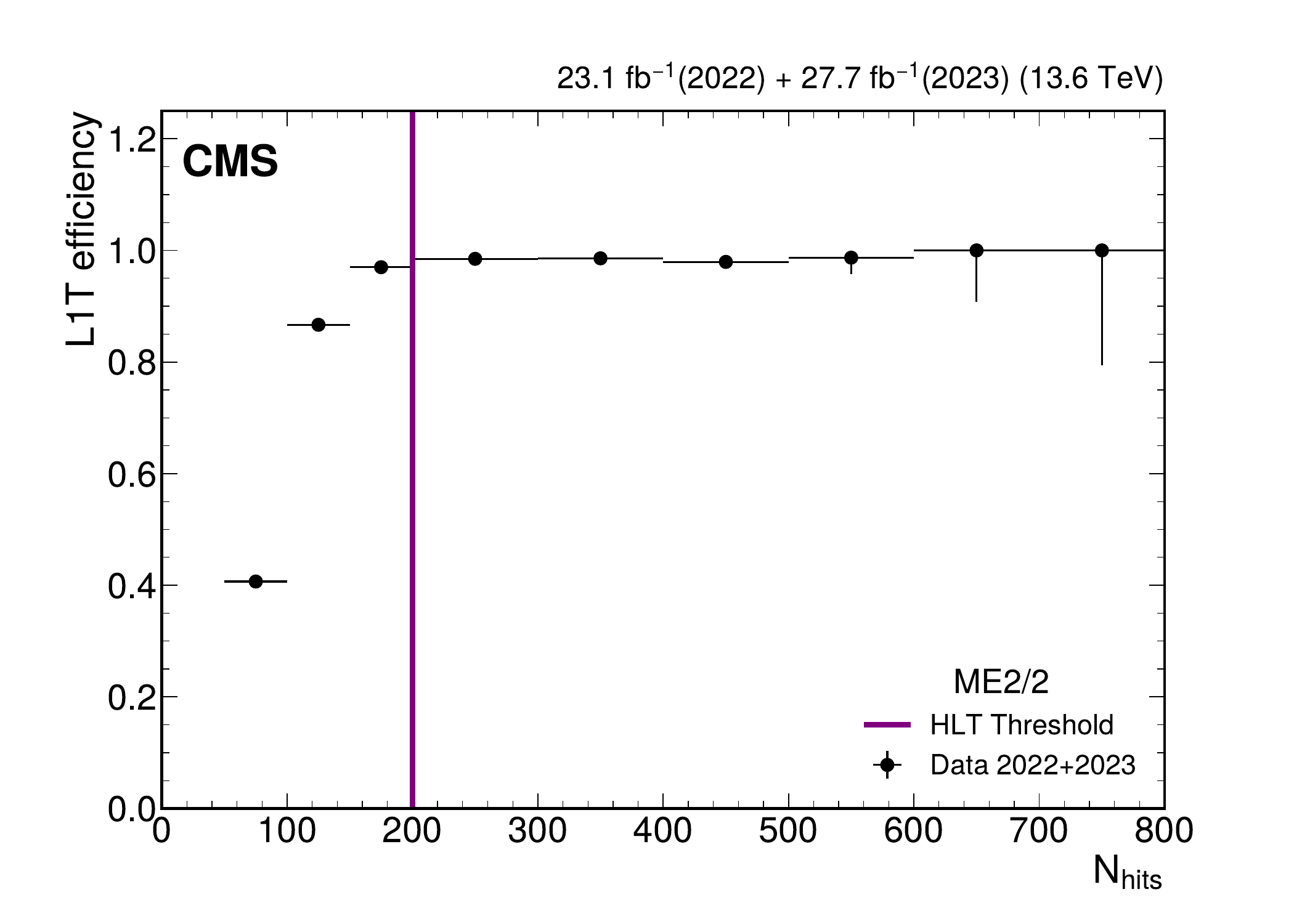}
\includegraphics[width=0.49\textwidth]{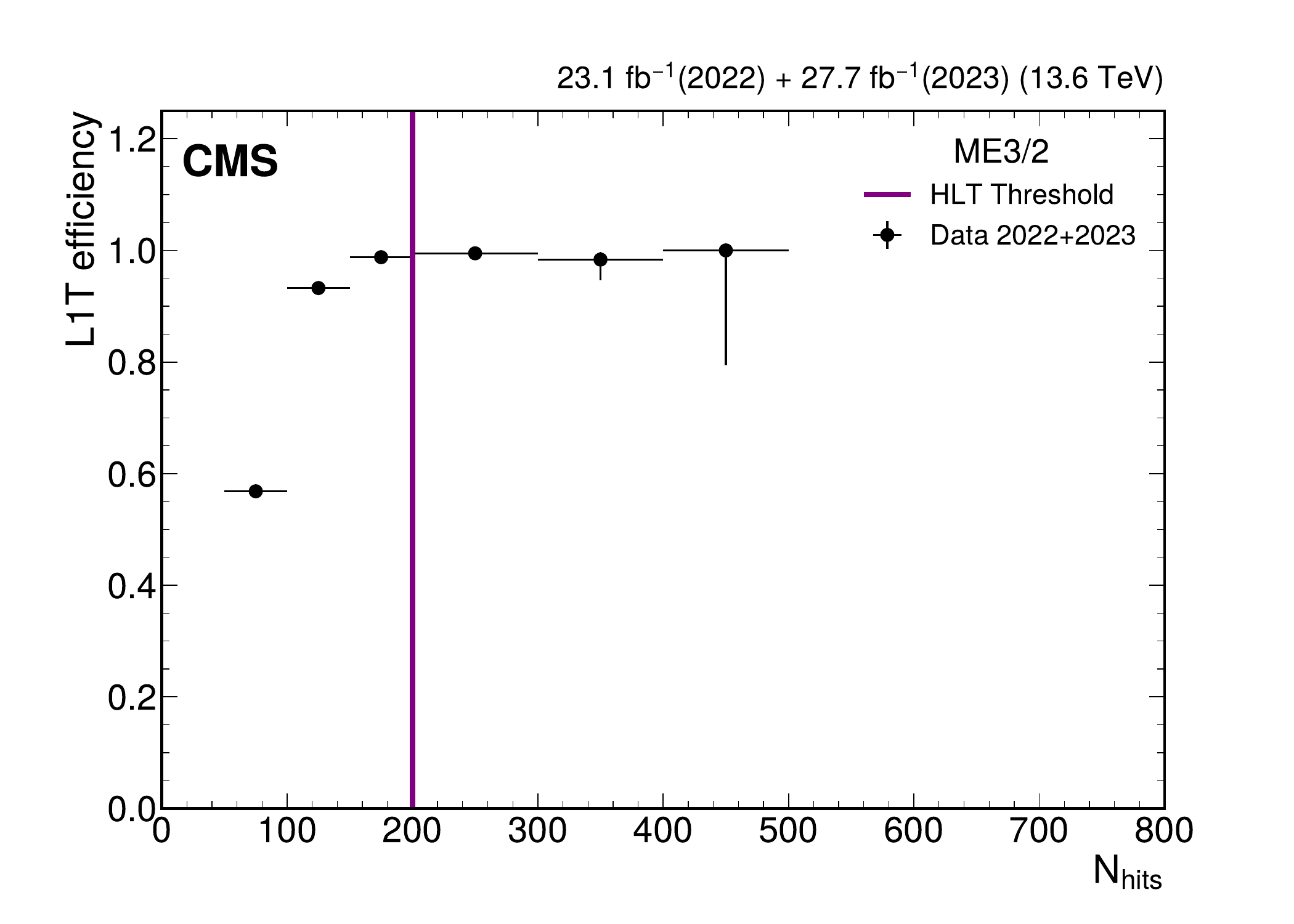}
\includegraphics[width=0.49\textwidth]{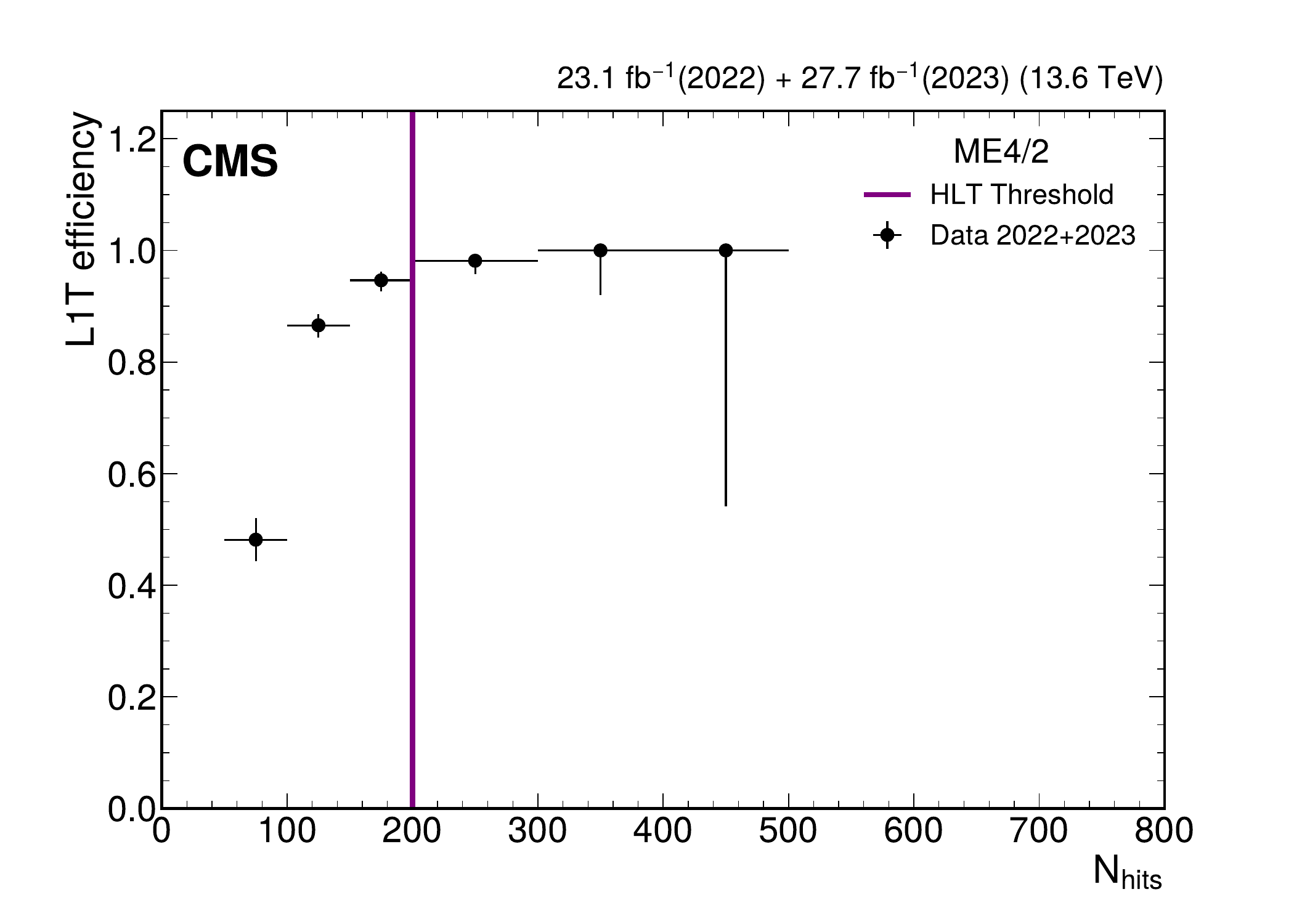}
\end{figure}

\begin{figure}
\caption{L1 trigger efficiencies of the One-Nominal trigger for inner ring chambers as a function of offline cluster size. Events are selected from the single-muon trigger and have exactly one cluster matched to a muon with $\Delta R <$ 0.4 , with at least 90\% of rechits in a single station-ring. \nhits refers to the number of offline CSC rechits after CA clustering with a distance parameter of 0.4. 
The turn-on feature is a result of the differences between the trigger object used in L1 and HLT (anode/cathode hit counts and HLT rechits respectively). Figures taken from Refs.~\cite{EXO-23-016,CMS-DP-2024-099}.}
\label{fig:inner_ring_eff}
\centering
\includegraphics[width=0.49\textwidth]{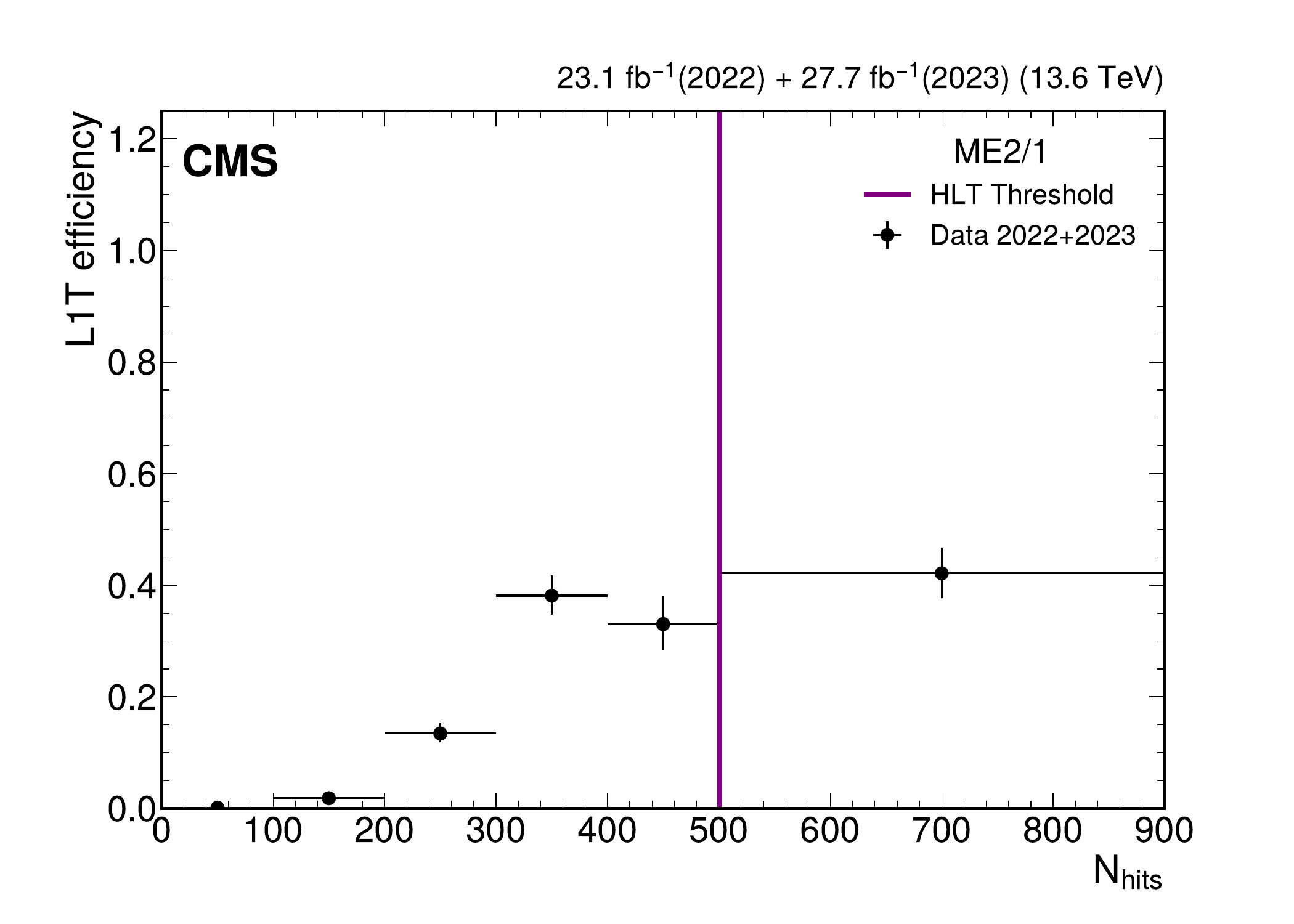}
\includegraphics[width=0.49\textwidth]{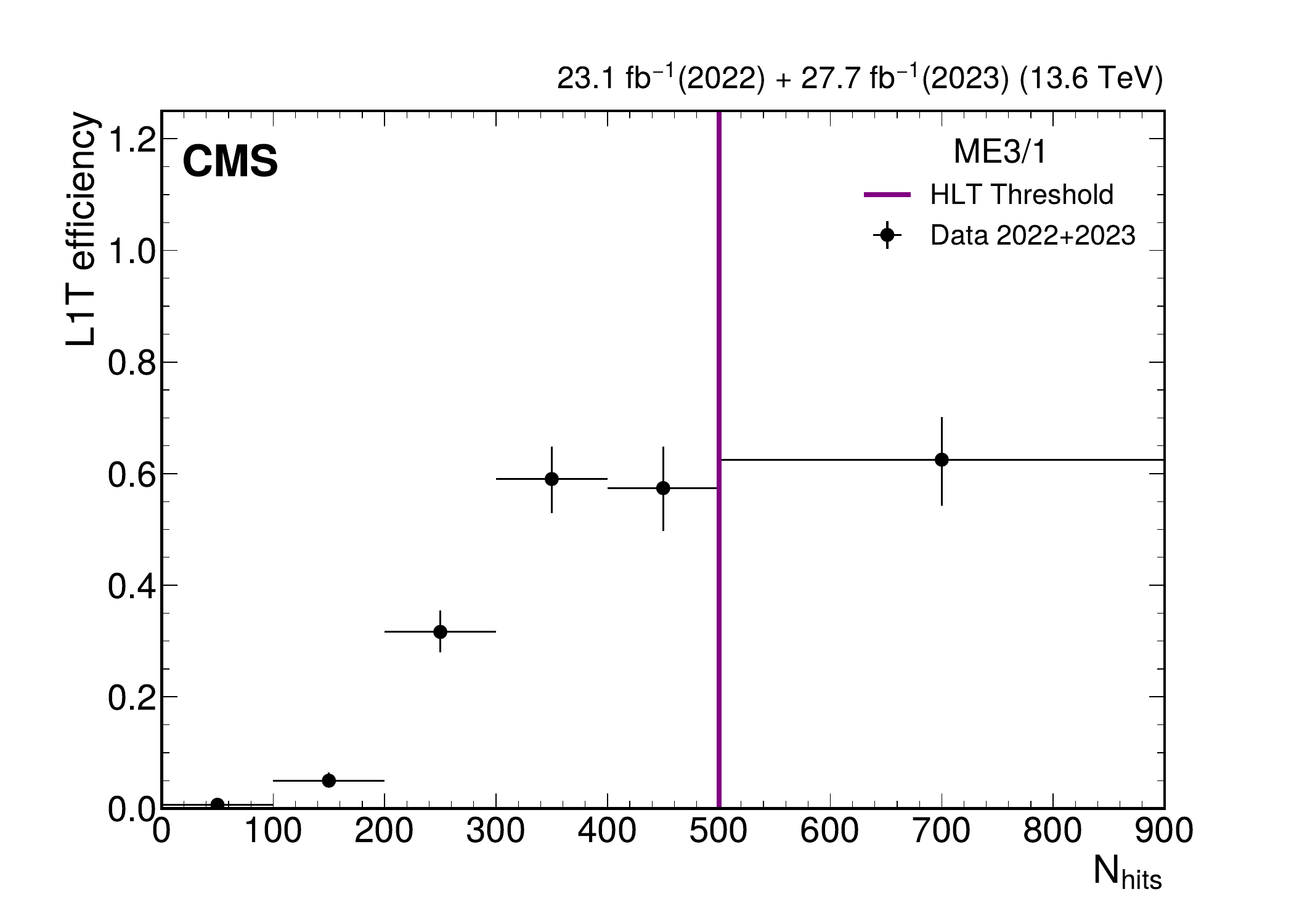}
\includegraphics[width=0.49\textwidth]{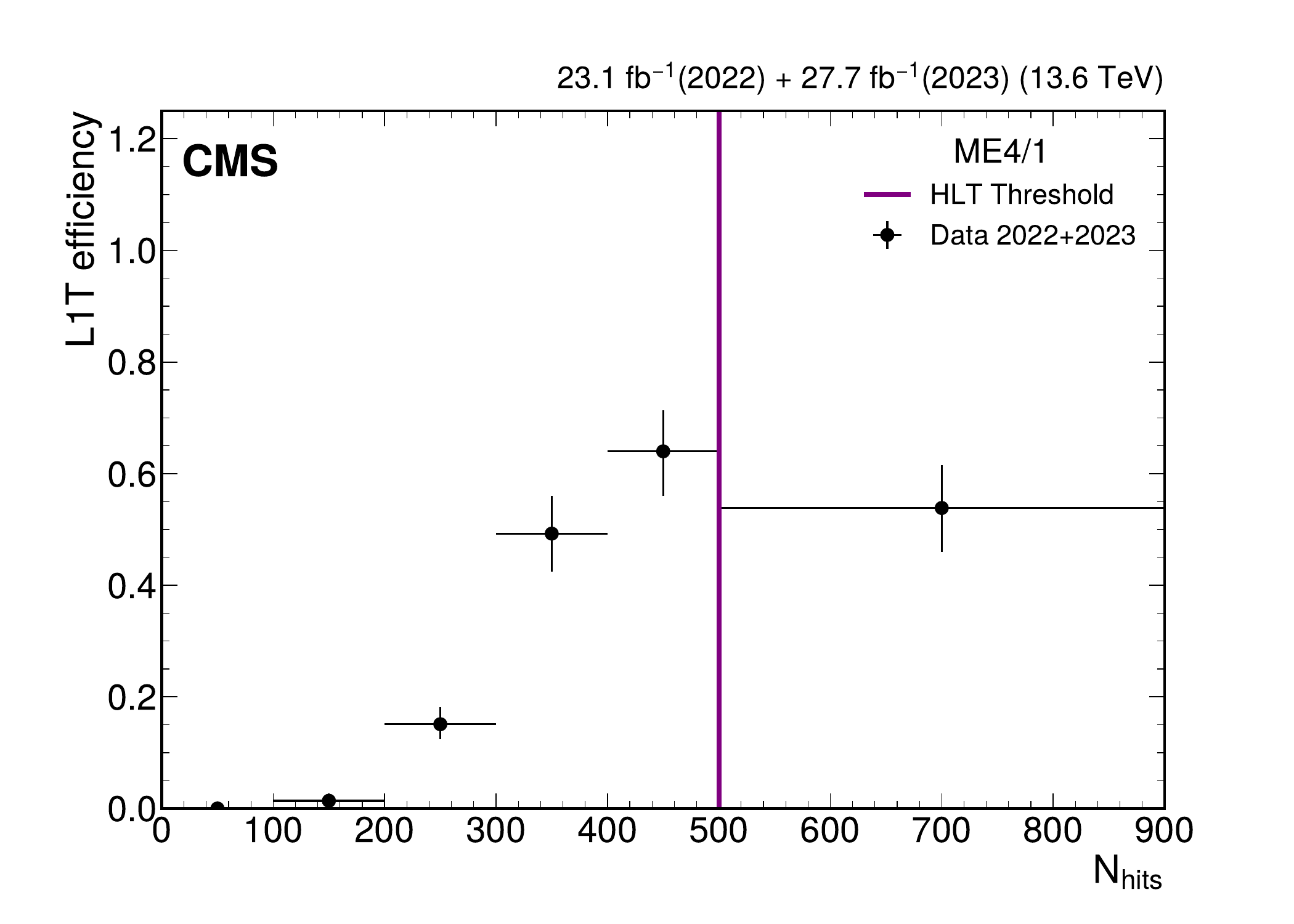}
\end{figure}

Because the trigger objects used in L1~(anode and cathode hits) and in HLT are different, we quantify the trigger efficiency at L1 with respect to HLT.
The trigger efficiency is measured by selecting clusters induced by muon bremsstrahlung as the proxy to the signal clusters, because there are limited collision processes in the standard model that can produce clusters in all muon stations.
Figures~\ref{fig:outer_ring_eff} and~\ref{fig:inner_ring_eff} show the trigger efficiency at the L1 level on a chamber-by-chamber basis for the outer ring and inner ring of the CSCs respectively. 
Events with clusters induced by muon brehmstrahlung are selected by a reference single-muon HLT algorithm that requires the presence of at least one isolated muon with $p_{T} >$ 24 GeV. 
A muon is required to pass loose identification criteria~\cite{muon_performance} and an additional set of criteria in $p_{T}$, $\eta$, and isolation for the cluster it induces to be considered for the efficiency measurement. 
We require the clusters to be geometrically matched to muons passing the chosen criteria in $\Delta R < 0.4$, defined as $\Delta R = \sqrt{\Delta\eta^{2} + \Delta\phi^{2}}$, where $\Delta\phi$ refers to the difference in the azimuthal angle between the two objects, and $\Delta\eta$ to the difference in $\eta$. 
The kinematic properties of a cluster are determined by averaging those of its constituent rechits. 

\par
The \nhits required to reach the plateau of L1 efficiencies is significantly higher for the outer ring chambers than the inner ring chambers due to the higher anode-wire count thresholds and the additional threshold on cathode-strip counts for the inner ring chambers.
The additional requirement for a minimum number of cathode hits to be in time with corresponding anode
hits for clusters in the inner ring of the CSC detector results in a decrease in efficiency for clusters with very large cluster size. 
\section{Future Developments\label{sec:future}}

The reach for some new physics signatures could significantly benefit from reducing shower thresholds at the trigger level, as this could lead to significant gains in signal efficiency. 
While the resulting increase in the trigger rate is prohibitive of doing this alone, for signatures like HNL, one can reduce the thresholds and keep the rate under control by requiring the presence of a charged lepton with moderate transverse momentum. 
This has been realized at the HLT level, but not at the L1 level. 
Two additional bits remain available at each stage of the L1 muon shower data transmission and allow for an implementation of such "loose" showers that can then be combined via a logical AND with a lepton requirement in the GT. 

On a longer timescale (targeting the HL-LHC), the upgraded L1 hardware~\cite{CERN-LHCC-2020-004} is designed to output 7.5 times higher trigger rate with around 3 times longer latency.
The upgraded hardware presents many opportunities for improving HMT and the physics reach.
First, HMT can be implemented in the CMS barrel with upgraded hardware in the DT system to provide more bandwidth and resources~\cite{CERN-LHCC-2017-012}, which could improve the acceptance to LLP models with more central decays by a factor of 2 to 3. 
Clustering would also be possible at the L1 level with the added bandwidth in the muon systems, which resembles the capability at the current HLT.
The Phase 2 L1 system will be capable of performing fast machine learning inference at the microsecond timescale, which is an opportunity to replace the simple clustering algorithm with a machine learning algorithm to exploit the finer details of signal showers, such as shape, depth, and potentially time information.

Another potential development can target searches for delayed objects with respect to the LHC collisions, such as jets or photons. 
If a trigger is deployed to select delayed clusters in the muon detectors, in contrast to the in-time clusters implemented in this paper, one can expect to extend the CMS Run-2 analysis sensitivity~\cite{delayedjets} to LLP displacements reaching the muon system. 
Muon detector hits in the neighboring bunch crossings of the triggered event (+1, +2 BXs) could be easily counted in the same way as for the in-time configuration illustrated in this paper.
Similarly, searches for stopped exotic LLPs~\cite{stopped_LLPs_2018} could benefit from the described trigger. 
This BSM model assumes that massive LLPs can be stopped inside the detector material if they lose all of their kinetic energy due to nuclear interactions or ionization. 
Thus, their decay products would be reconstructed in delayed events unrelated to their production if the stopped LLPs have long lifetimes. 
The Beam Pick-up Timing for eXperiments (BPTX) detectors~\cite{4774822}, which consist of standard LHC button electrodes that measure the position, timing, and intensity of the beams, can be used to select out-of-coincidence signals in the muon detectors with any of the pp collisions, and thresholds on out-of-time hits could be set for triggering.
 
\section{Summary\label{sec:summary}}

We have reported on the first implementation of a high multiplicity trigger (HMT) specifically designed for detecting the signatures of long-lived particles (LLP) decaying in the CMS muon system.
Such LLP signatures arise in a wide range of highly motivated models of physics beyond the Standard Model (BSM), and represents one of the most active experimental frontiers in searches for new physics at the Large Hadron Collider.
The newly implemented HMT trigger not only represent the first LLP-specific hardware triggers for the CMS experiment, but also opens up a whole new program of LLP searches using the muon detector shower experimental signature sensitive to a very wide array of BSM models with macroscopically long lifetimes and decay lengths.

We employed a simple and highly efficient algorithm for the Cathode Strip Chamber (CSC) detectors of the CMS experiment, based on the multiplicity of anode wire and cathode strip hits per chamber. The presence of a shower signature is encoded in 2-bits, allowing decisions to be made based on multiple levels of thresholds. The trigger multiplicity thresholds were optimized separately per station, taking into account the varying rate of background in different stations.
The HMT improves signal efficiency by over a factor of 30 compared to previous large-\pt-imbalance triggers for benchmark LLP models.
We evaluated the efficiency of the trigger using showers from muons undergoing bremsstrahlung in dimuon events produced by the Drell-Yan process and demonstrated near 100\% level-1 trigger efficiency above the High-Level Trigger (HLT) threshold for the outer ring stations, where the background event rate is lower.
Finally, we demonstrated linear dependence of the trigger rate on the amount of pileup.

Future improvements and follow-up work include reducing thresholds and improving acceptance by leveraging on the presence of other objects for associated production of LLP’s and importantly implementing a similar trigger for the barrel muon detectors enabled by upgraded hardware from the CMS Phase 2 upgrade.

\section{Acknowledgements}
We thank Michele Papucci for insightful discussions on the impact of the High Multiplicity Trigger on a wide variety of BSM models.

We congratulate our colleagues in the CERN accelerator departments for the excellent performance of the LHC and thank the technical and administrative staffs at CERN and at other CMS institutes for their contributions to the success of the CMS effort. In addition, we gratefully acknowledge the computing centres and personnel of the Worldwide LHC Computing Grid and other centres for delivering so effectively the computing infrastructure essential to our analyses. Finally, we acknowledge the enduring support for the construction and operation of the LHC, the CMS detector, and the supporting computing infrastructure provided by the following funding agencies: SC (Armenia), BMBWF and FWF (Austria); FNRS and FWO (Belgium); CNPq, CAPES, FAPERJ, FAPERGS, and FAPESP (Brazil); MES and BNSF (Bulgaria); CERN; CAS, MoST, and NSFC (China); MINCIENCIAS (Colombia); MSES and CSF (Croatia); RIF (Cyprus); SENESCYT (Ecuador); ERC PRG and PSG, TARISTU24-TK10 and MoER TK202 (Estonia); Academy of Finland, MEC, and HIP (Finland); CEA and CNRS/IN2P3 (France); SRNSF (Georgia); BMFTR, DFG, and HGF (Germany); GSRI (Greece); MATE and NKFIH (Hungary); DAE and DST (India); IPM (Iran); SFI (Ireland); INFN (Italy); MSIT and NRF (Republic of Korea); MES (Latvia); LMTLT (Lithuania); MOE and UM (Malaysia); BUAP, CINVESTAV, CONACYT, LNS, SEP, and UASLP-FAI (Mexico); MOS (Montenegro); MBIE (New Zealand); PAEC (Pakistan); MSHE, NSC, and NAWA (Poland); FCT (Portugal); MESTD (Serbia); MICIU/AEI and PCTI (Spain); MOSTR (Sri Lanka); Swiss Funding Agencies (Switzerland); MST (Taipei); MHESI (Thailand); TUBITAK and TENMAK (T\"{u}rkiye); NASU (Ukraine); STFC (United Kingdom); DOE and NSF (USA).

\bibliographystyle{JHEP}
\bibliography{references}
\end{document}